\documentclass[fleqn,usenatbib]{mnras}

\usepackage{newtxtext,newtxmath}
\usepackage[T1]{fontenc}

\DeclareRobustCommand{\VAN}[3]{#2}
\let\VANthebibliography\thebibliography
\def\thebibliography{\DeclareRobustCommand{\VAN}[3]{##3}\VANthebibliography}

\usepackage{longtable}
\usepackage{pdflscape}
\usepackage{afterpage}
\usepackage{adjustbox}
\usepackage{placeins}
\usepackage{capt-of}
\usepackage{textcase}
\usepackage{caption}
\usepackage{array}
\usepackage{natbib}
\usepackage{multirow}
\usepackage{tabularx}
\usepackage{booktabs}
\usepackage{tablefootnote}
\usepackage{threeparttable} % to use table notes
\usepackage{graphicx}	% Including figure files
\usepackage{amsmath}
\title[Fullerenes in the circumstellar medium of Herbig Ae/Be stars]{Fullerenes in the circumstellar medium of Herbig Ae/Be stars: Insights from the Spitzer mid-infrared spectral catalog}

\author[R. Arun et al.]{
R. Arun$^{1,2}$\thanks{E-mail: arunroyon@gmail.com},
Blesson Mathew$^{2}$,
P. Manoj$^{3}$,
G. Maheswar$^{1}$,
B. Shridharan$^{2}$,
Sreeja S. Kartha$^{2}$, and
\newauthor
Mayank Narang$^{3,4}$
\\
$^{1}$Indian Institute of Astrophysics, Sarjapur Road, Koramangala, Bangalore 560034, India\\
$^{2}$Department of Physics and Electronics, CHRIST (Deemed to be University), Bangalore 560029, India\\
$^{3}$Tata Institute of Fundamental Research, Homi Bhabha Road, Mumbai 400005, India\\
$^{4}$Institute of Astronomy and Astrophysics, Academia Sinica, No. 1, Section 4, Roosevelt Road, Taipei 10617, Taiwan
}
\date{Accepted 2023 May 15. Received 2023 May 15; in original form 2023 March 29}

\pubyear{2023}

\begin{document}
\label{firstpage}
\pagerange{\pageref{firstpage}--\pageref{lastpage}}
\maketitle

% Abstract of the paper
\begin{abstract}
 This study presents the largest mid-infrared spectral catalog of Herbig Ae/Be stars to date, containing the Spitzer Infrared Spectrograph spectra of 126 stars. Based on the catalog analysis, two prominent infrared vibrational modes of C\textsubscript{60} bands at 17.4 $\mu m$ and 18.9 $\mu m$ are detected in the spectra of nine sources, while 7.0 $\mu m$ feature is identified in the spectra of HD 319896. The spectral index analysis and the comparison of the known sources with C\textsubscript{60} features indicated that there exist two different types of emission classes among the sample of stars. The infrared spectra of six Herbig Ae/Be stars in this study resemble that of reflection nebulae, and their association with previously known reflection nebulae is confirmed. In the case of three Herbig Ae/Be stars we report the tentative evidence of C\textsubscript{60} emission features originating from the circumstellar disk or nearby diffused emission region. The detection fraction of C\textsubscript{60} in the total HAeBe star sample is $\sim$ 7\%, whereas the detection fraction is 30\% for HAeBe stars associated with nebulosity. In the catalog, C\textsubscript{60} is exclusively present in the circumstellar regions of B type Herbig Ae/Be stars, with no evidence of its presence detected in stars with later spectral types. The present study has increased the number of young stellar objects and reflection nebulae detected with C\textsubscript{60} multifold, which can help in understanding the excitation and formation pathway of the species.
\end{abstract}

% Select between one and six entries from the list of approved keywords.
% Don't make up new ones.
\begin{keywords}
stars: variables: Herbig Ae/Be -- astrochemistry -- infrared: general -- ISM: molecules
\end{keywords}

%%%%%%%%%%%%%%%%%%%%%%%%%%%%%%%%%%%%%%%%%%%%%%%%%%

%%%%%%%%%%%%%%%%% BODY OF PAPER %%%%%%%%%%%%%%%%%%

\section{Introduction}

Studying young stellar objects (YSOs) is crucial for understanding the formation and evolution of stars and their environments. Herbig Ae/Be (HAeBe) stars, in particular, are of great interest as they represent a "missing link" between low-mass T Tauri stars and Massive YSOs and possess circumstellar accretion disks \citep{WATERS1998}. Emission lines in the spectra of HAeBe stars reveal the dynamics and state of the gaseous component in the circumstellar disk. At the same time, their spectral energy distribution (SED) shows an infrared (IR) excess indicative of hot and/or cool dust in the circumstellar medium \citep{Hillenbrand1992, Malfait1998}. The mid-IR (MIR) spectrum of HAeBe stars has revealed the presence of polycyclic aromatic hydrocarbon (PAH; \citealp{Brooke1993AJ....106..656B,Meeus2001A&A...365..476M}) molecules in their circumstellar region. The largest sample (55) of HAeBe stars studied as a class of objects is by \citet{Seok2017ApJ...835..291S}, where they found that 70\% of HAeBe stars are associated with PAH emission. Most MIR spectral studies of HAeBe stars have focused on the analysis of PAH features. A comprehensive compilation of the MIR spectra of HAeBe stars and the analysis of other molecular features is currently lacking in the literature.
\begin{figure*}

    \includegraphics[width=2\columnwidth]{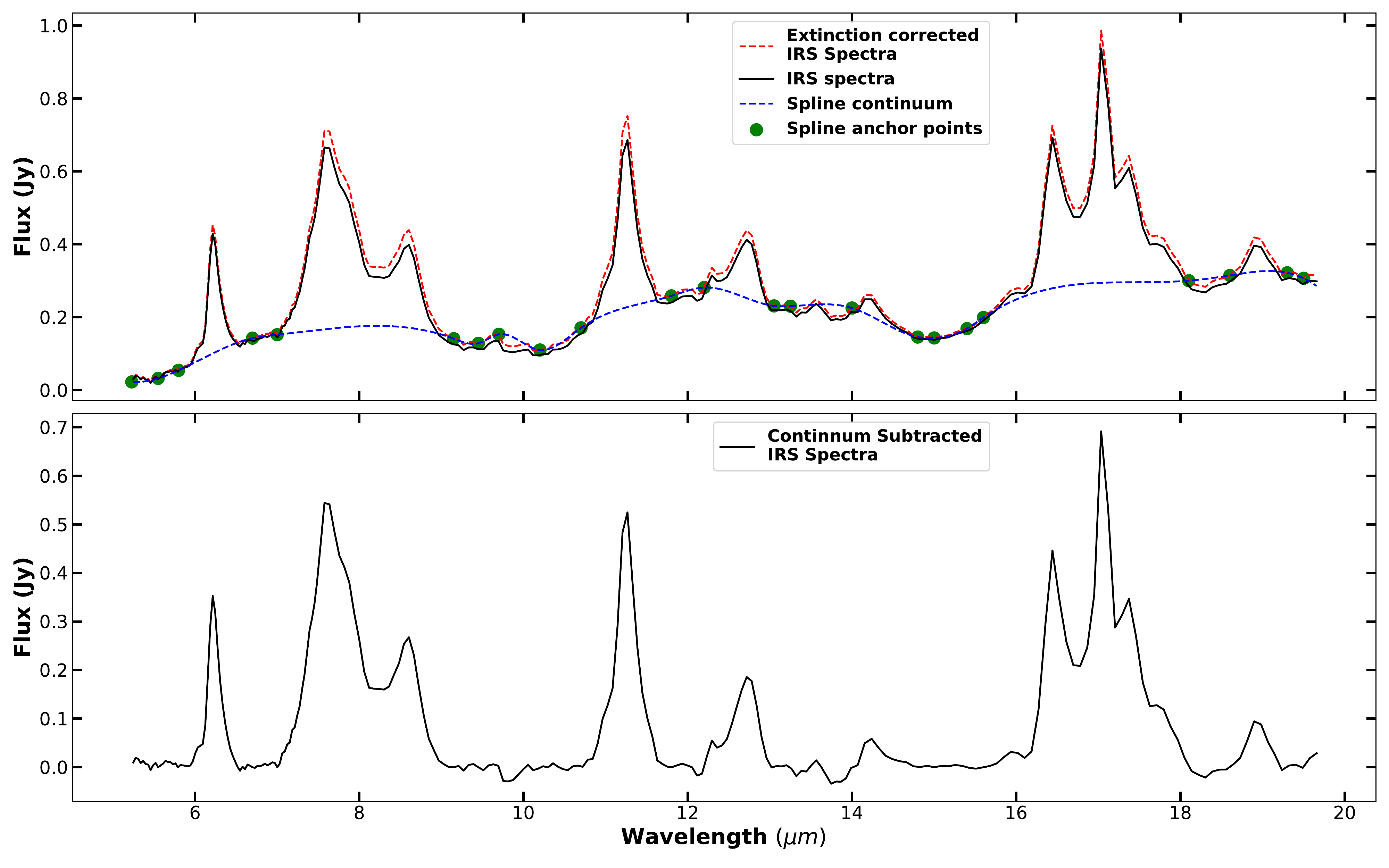}
 
    \caption{The figure shows the representative example of extinction correction and continuum subtraction of IRS spectra. The uncorrected IRS spectrum of HBC 334 is shown in black and the dereddened spectrum is shown in red (top panel). The spline continuum and its anchor points are also shown in the figure. The dereddened and continuum subtracted spectrum is shown in the bottom panel.}
    \label{fig:1st}
\end{figure*}
To address this gap, we leverage the large archive of reduced Infrared Spectrograph (IRS; \citealp{Houck2004ApJS..154...18H}) spectra from the Spitzer Space Telescope \citep{Werner2004ApJS..154....1W}, named as Combined Atlas of Sources with Spitzer IRS Spectra (CASSIS\footnote{CASSIS is a product of the IRS instrument team, supported by NASA and JPL. CASSIS is supported by the``Programme National de Physique Stellaire" (PNPS) of CNRS/INSU co-funded by CEA and CNES and through the ``Programme National Physique et Chimie du Milieu Interstellaire" (PCMI) of CNRS/INSU with INC/INP co-funded by CEA and CNES.}), to create a catalog of HAeBe stars. Though some of the previous studies have concentrated on Herbig Ae (HAe) stars \citep{Acke2010}, we attempt to do a comparative study of HAe and HBe MIR spectra through this work and may facilitate the detection of other molecular species. Ultimately, this catalog will contribute to advancing our understanding of the physical processes and chemical composition of the circumstellar environments of HAeBe stars and can act as a reference for future high-resolution observations with James Webb Space Telescope ({\it JWST}).

Buckminsterfullerene, C\textsubscript{60}, is one of the most stable cage-like carbonaceous molecules, which is postulated to be ubiquitous in space \citep{Kroto1985Natur.318..162K}. The electronic spectrum of C\textsubscript{60} has three strong broad peaks at 216, 264, and 339 $nm$ and four infrared (IR) vibrational modes at 7.0, 8.5, 17.4, and 18.9 $\mu m$ (e.g., \citealp{Zhang2013EP&S...65.1069Z}). \cite{Sellgren2007ApJ...659.1338S} detected the 17.4 and 18.9 $\mu m$ features towards reflection nebulae (RNe) NGC 7023 and has discussed the possibility that these features are due to C\textsubscript{60}. It was later confirmed by detecting the third C\textsubscript{60} emission feature at 7.04 $\mu m$ in NGC 7023 and detection in second RNe, NGC 2023 \citep{Sellgren2010ApJ...722L..54S}. The first observational evidence for all the four vibrational modes of C\textsubscript{60} in space is confirmed by its detection in the planetary nebula (PNe) Tc 1 \citep{Cami2010Sci...329.1180C}. Follow-up studies identified C\textsubscript{60} in numerous Galactic and extragalactic PNe sources \citep{Garc2010ApJ...724L..39G, Garc2011ApJ...737L..30G, Otsuka2014MNRAS.437.2577O}. C\textsubscript{60} has also been identified in variety of environments such as diffuse ISM \citep{Berne2017A&A...605L...1B}, and star-forming regions \citep{Rubin2011MNRAS.410.1320R, Iglesias2019MNRAS.489.1509I}. The C\textsubscript{60} molecule is reported in only a few YSOs \citep{Roberts2012MNRAS.421.3277R,Iglesias2019MNRAS.489.1509I}. Interestingly, there is only one instance of detecting C\textsubscript{60} near a HAeBe star, i.e., in the case of HD 97300 \citep{Roberts2012MNRAS.421.3277R}.

One of the defining criteria for identifying HAeBe stars is its association with nebulosity \citep{Herbig1960ApJS....4..337H}. It can be seen that some of the well-studied RNe are found to be associated with HAeBe stars. For example, RNe NGC 7023, which is the first RNe reported to show C\textsubscript{60} emission features in the IRS spectrum, is associated with the well-known HAeBe star HD 200775 \citep{Sellgren2007ApJ...659.1338S,Sellgren2010ApJ...722L..54S}. Similarly, the first HAeBe star identified with C\textsubscript{60}, HD 97300, is associated with RNe IC 2631 \citep{Magakian2003A&A...399..141M,Roberts2012MNRAS.421.3277R}. Even though the association of nebulosity with HAeBe stars is not universal, the environments of HAeBe stars are ideal regions to survey for C\textsubscript{60} emission features. Also, no previous studies were performed to search for C\textsubscript{60} near the environment of HAeBe stars. The study of another class of objects with C\textsubscript{60} emission features other than PNe can help understand their formation mechanism and helps to evaluate their role as carriers for spectral features such as diffuse interstellar bands. Furthermore, nanodiamonds have been detected in the circumstellar medium of HAeBe stars \citep{Guillois1999ApJ...521L.133G}. \cite{Goto2009ApJ...693..610G} proposed the formation of nanodiamonds in the environment of fullerene-type molecules, under the influence of high energetic shocks. Hence, by identifying a sample of HAeBe stars with C\textsubscript{60} emission features, one can also address the possible formation channels of nanodiamonds.

In this study, we present the creation of a comprehensive catalog of 126 HAeBe stars observed with the Spitzer IRS instrument, making it the largest MIR catalog to date. Using these observations, we report the detection of C\textsubscript{60} emission features in nine HAeBe stars. The sample selection process is described in Section 2. We present our results in Section 3, which includes details of the C\textsubscript{60} detection and the likely origin of the C\textsubscript{60} features in the vicinity of HAeBe stars. In Section 4, we briefly discuss the implications of our results for the formation mechanism of C\textsubscript{60} near HAeBe stars. Finally, we summarize our results in Section 5.

\section{A MIR spectral catalog of HAeBe stars}

The present study aims to identify C\textsubscript{60} near HAeBe stars. We have analyzed a sample of HAeBe stars selected from a well-established catalog using MIR spectra obtained from the Spitzer IRS. This section will describe the methodology employed for the sample selection and outline the observational data utilized in our analysis.

\begin{table*}
    \begin{center}
    \caption{The stellar parameters and the spectral information of all the 7 HAeBe stars identified with C\textsubscript{60} features. The distance estimates are taken from  \protect \cite{Bailer-Jones2021AJ....161..147B}. The age and mass estimates are from  \protect \cite{Vioque2018A&A...620A.128V}. Information on RNe association is taken from  \protect \cite{Magakian2003A&A...399..141M}.}
    \label{tab:full_table}
    \begin{tabular}{cccccccccc}
    \hline
    Source & AORkey & Spectral Type & Distance (pc) & Age (Myr) & Mass (M\textsubscript{\(\odot\)}) & PAHs & 6--9 $\mu m$ plateau & 17 $\mu m$ plateau\\
     \hline
        \textit{Sources with associated RNe} \\
        CPM 25 & 25736192 & B2 & 3415 $\pm$ 575 & $0.7^{+5.7}_{-0.4}$ & $5.2^{+2.2}_{-1.2}$ & Yes & No & Yes \\          
        BD+30 549\textsuperscript{*} & 14121472 & B8 & 285 $\pm$ 2 & $5.48^{+15}_{-2}$ & $2.28^{+0.37}_{-0.19}$ & Yes & No & Yes \\ 
        HBC 334 & 25731328 & B3 & 1563 $\pm$ 55 & $2.1^{+4.3}_{-1.1}$ & $2.1^{+4.3}_{-1.1}$ & Yes & No & Yes \\
        PDS 344 & 25738241 & B5 & 2158 $\pm$ 52 & $1.8^{+8.4}_{-0.2}$ & $3.48^{+0.17}_{-0.23}$ & Yes & No & Yes \\ 
        LkHa 215 & 14124032 & B7 & 725 $\pm$ 14 & $1.03^{+8.4}_{-0.2}$ & $3.8^{+0.59}_{-0.36}$ & Yes & No & Yes \\ 
        HD 46060 & 25732864 & B3-B4 & 932.9 $\pm$ 82 & $0.09^{+0.01}_{-0.1}$ & $9.6^{+3.4}_{-2.4}$ & No & No & Yes \\ 
    \hline
    \textit{Sources without associated RNe}\\
        MWC 593 & 25746432 & B4 & 1341.4 $\pm$ 180 & $0.12^{+0.08}_{-0.05}$ & $8.0^{+1.8}_{-1.2}$ & No & No & No \\ 
        HD 319896 & 25739776 & B4 & 1285 $\pm$ 46 & $0.3^{+0.18}_{-0.13}$ & $5.9^{+1.2}_{-0.8}$ & No & Yes & No \\
        SAO 220669 & 25737216 & B4 & 870 $\pm$ 9 & $0.13^{+0.08}_{-0.05}$ & $7.9^{+1.2}_{-1.1}$ & No & Yes & No \\
 \hline
    \end{tabular}
    \raggedright \\(*)- BD+30 549 has three observations in the 20\arcsec~radius.\\
     \end{center}
\end{table*}

The coordinates of 252 confirmed HAeBe stars were obtained from the recent catalog of \cite{Vioque2018A&A...620A.128V}. A search was carried out in CASSIS archive with a search radius of 20\arcsec, resulting in the identification of low-resolution IRS (SL/LL) spectra for 94 stars, with a wavelength span of 5--38 $\mu m$ for 55 stars and 5--15 $\mu m$ for the remaining 39 stars. High-resolution IRS (SH/LH) spectra were identified for 79 stars, with 47 stars having both low and high-resolution spectra. This represents the largest (126) collection of Spitzer IRS spectra of HAeBe stars  to date; a table with relevant details is given in the appendix.

CASSIS database is an archive of high-quality spectra containing the IRS low-resolution and high-resolution (staring-mode) spectra taken during the Spitzer mission. SMART-AdOpt, an automated spectrum extraction tool that can do optimal (as well as regular) extractions utilising a super-sampled point spread function (PSF) to produce the best possible signal-to-noise ratio, is used by CASSIS for low-resolution spectral reduction \citep{CASSIS2011ApJS..196....8L}. Three extraction methods exist for CASSIS low-resolution spectral reduction (optimal, tapered column and full slit). The ideal extraction for point-like sources is optimal extraction, in which the PSF is utilised to scale the source's spatial profile. A tapered column extraction is ideal for sources that have been partly extended since it recovers most of the source's flux. A full slit extraction is better suited for very extended sources. CASSIS has updated its extraction routines consistently, and the current version of low-resolution spectra is LR7 (a detailed description of the extraction process is available on the CASSIS website\footnote{https://irs.sirtf.com/Smart/CassisLRPipeline}). CASSIS also use super-sampled PSF (empirical) to extract high-resolution IRS spectra. For the high-resolution spectral reduction, two extraction methods are employed - optimal extraction for point-like sources and full-aperture extraction for extended objects \citep{Lebouteiller2015ApJS..218...21L}. The current version for high-resolution spectra is HR1 (detailed extraction procedure is given in the website\footnote{https://irs.sirtf.com/Smart/CassisHRPipeline}).

The low-resolution IRS spectra (SL/LL) has a resolving power (R) of 64 - 128 \citep{Houck2004ApJS..154...18H}. Most of the C\textsubscript{60} identification studies have used the high-resolution mode (LH), with R = 600 \citep{Cami2010Sci...329.1180C, Roberts2012MNRAS.421.3277R}. However, C\textsubscript{60} emission features are also identified using the low-resolution mode in the studies of Galactic PNe K3-54, SMC PNe SMP SMC 16 \citep{Garc2010ApJ...724L..39G} and the star-forming region IC 348 of the Perseus molecular cloud \citep{Iglesias2019MNRAS.489.1509I}. This implies that the 18.9 $\mu m$ C\textsubscript{60} feature is also detectable using the low-resolution IRS instrument. Therefore, we consider both low and high-resolution IRS spectra of HAeBe stars in the search for the C\textsubscript{60} features.

The spatial extent of the source automatically decides the best extraction of low-resolution spectra. It could be either optimal, tapered column or full slit extraction. We performed the CASSIS-recommended extraction for all 94 sources for the initial survey. There are stars with multiple spectra in the 20\arcsec~search radius. In those cases, we take the closest spectra to the stellar coordinates. In the HAeBe star BD +30 549, multiple spectra are available with increasing separation from the source. This particular source and its IRS spectra were used in PAH evolution studies \citep{Andrews2015ApJ...807...99A,Boersma2016ApJ...832...51B}. C\textsubscript{60} has four infrared vibrational modes at 7.0, 8.5, 17.4, and 18.9 $\mu m$, which can be seen using the {\it Spitzer} IRS spectra. The three features at 7.0, 8.5, and 17.4 $\mu m$ are usually blended with intense PAH features. The 18.9 $\mu m$ C\textsubscript{60} feature is not affected due to the transitions from other molecular species or PAH emission features. So the current sample consists of 55 low-resolution spectra, which includes the 18.9 $\mu m$ feature and 80 high-resolution spectra of HAeBe stars.

The IRS spectra of all HAeBe stars are extinction corrected before further analysis. We adopted the composite extinction law prescribed by \cite{McClure2009ApJ...693L..81M} for the extinction correction. This method has been successfully used to deredden the IRS spectra of young stars in the nearby young clusters and star-forming regions \citep{Furlan2009ApJ...703.1964F, McClure2010ApJS..188...75M, Manoj2011ApJS..193...11M}. The steps to deredden the IRS spectra are as follows: objects with A$_{J}$ $<$ 0.8 mag are dereddened using extinction law from \cite{Mathis1990ARA&A..28...37M} with  total-to-selective extinction ratio (R\textsubscript{V} = 5). For objects with A$_{J}$ $>$ 0.8 mag, we used the appropriate extinction curves from \cite{McClure2009ApJ...693L..81M} to deredden the IRS spectra. A representative example of extinction correction is shown in the \autoref{fig:1st}(top panel). A higher value of R\textsubscript{V} is justified because HAeBe stars are reported having the grain growth in the disk \citep{Gorti1993A&A...270..426G}. Also, R\textsubscript{V} = 5 is used in the numerous HAeBe star studies \citep{Hernandez2004AJ....127.1682H,manoj2006,Mathew2018ApJ...857...30M,Arun2019AJ....157..159A}. The extinction curves of \cite{McClure2009ApJ...693L..81M} is used in the recent infrared studies of HAeBe stars \citep{Zhang2022ApJS..259...38Z,Grant2022ApJ...926..229G}.

The continuum subtraction process for IRS spectra of HAeBe stars was carried out using spline fitting with predefined anchor points. The anchor points for the 5-15 $\mu$m region were adopted from \cite{Seok2017ApJ...835..291S}. For spectra dominated by PAHs, we utilized anchor points at specific wavelengths of 5.55, 5.80, 6.7, 7.0, 9.15, 9.45, 9.7, 10.2, 10.7, 11.8, 12.2, 13.05, 13.25, 13.8, 14.0, and 14.8 $\mu$m, as illustrated in \autoref{fig:1st}. Slightly different anchor points were used for IRS spectra dominated by silicate emission. We used anchor points at 7.35, 8.95, and 9.35 $\mu$m instead of 9.15 and 9.45 $\mu$m, while keeping the other points the same as those for PAH-dominated sources. Anchor points for the 15-20 $\mu$m region were taken from \cite{Acke2010}, where we used points at 15.0, 15.4, 15.6, 18.1, 18.6, 19.3, and 19.5 $\mu$m. An example of continuum-subtracted spectra is shown in \autoref{fig:1st} (bottom panel). Interestingly, three sources that are identified without PAHs but with C\textsubscript{60} (discussed in Sect. 3.1). These stars are given modified anchor points of 5.45, 6.0, 9.15, 9.45, 9.7, 10.2, 10.7, 11.8, 12.2, 13.05, 13.25, 13.8,15.3, 16.,16.9, 17.7,18.5, 19.3, and 19.5 $\mu$m. A subset of points are used in the case of high-resolution sources. Finally, we created extinction-corrected and continuum-subtracted IRS spectra for 126 HAeBe stars.

%The IRS spectra of 126 HAeBe stars were extinction corrected and continuum subtracted for visual inspection to identify C\textsubscript{60} molecular emission.

\section{Results}
 
\begin{figure*}

    \includegraphics[width=2\columnwidth]{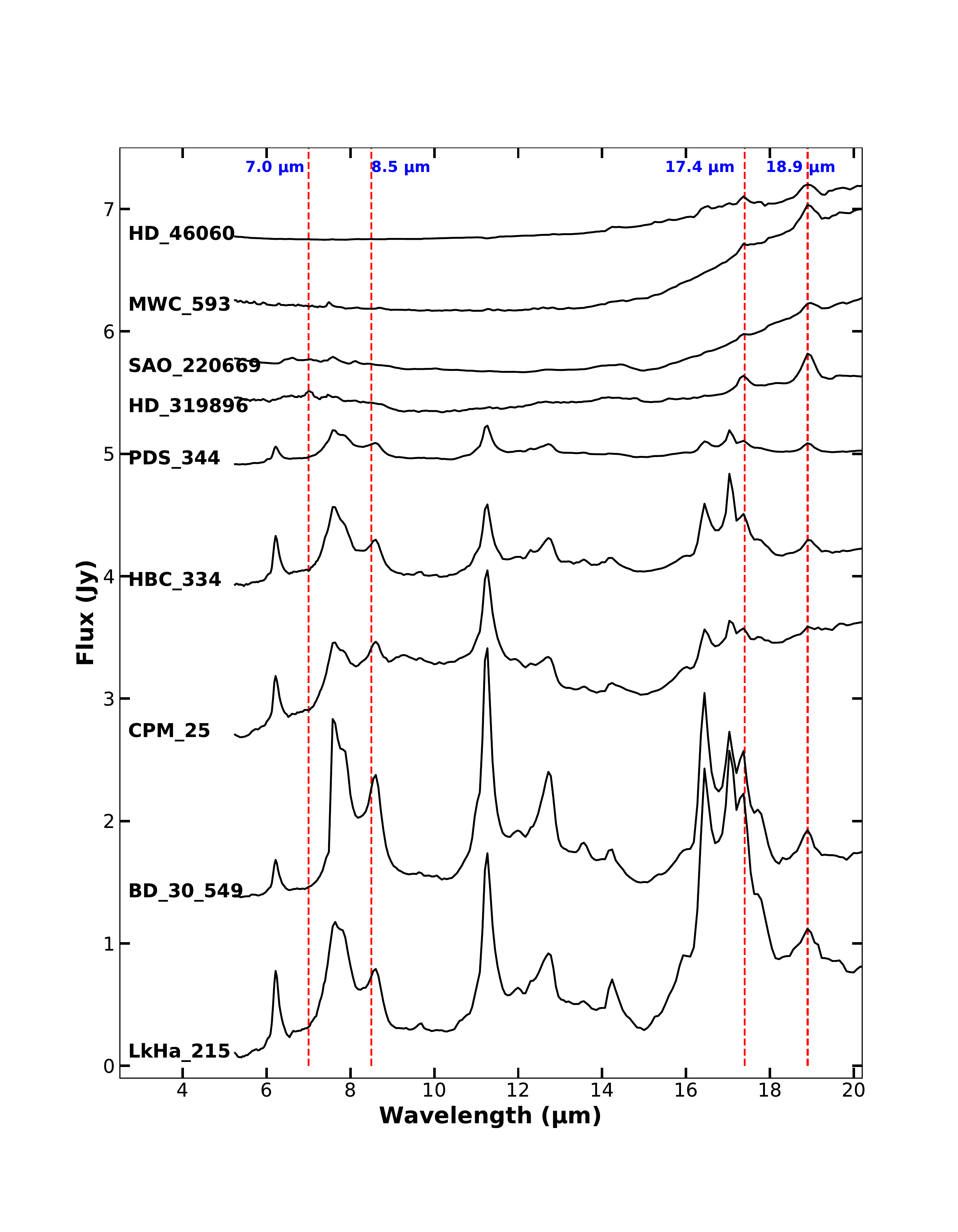}
 
    \caption{The figure shows the low-resolution Spitzer IRS spectra of seven HAeBe stars with C\textsubscript{60} emission features. The flux values are given an offset for better visualization of the spectra. The four vibrational modes of C\textsubscript{60} are shown as red lines. The LL region is normalized to the SL mode flux for visualization.}
    \label{fig:1}
\end{figure*}
\begin{figure*}
    \includegraphics[width=2\columnwidth]{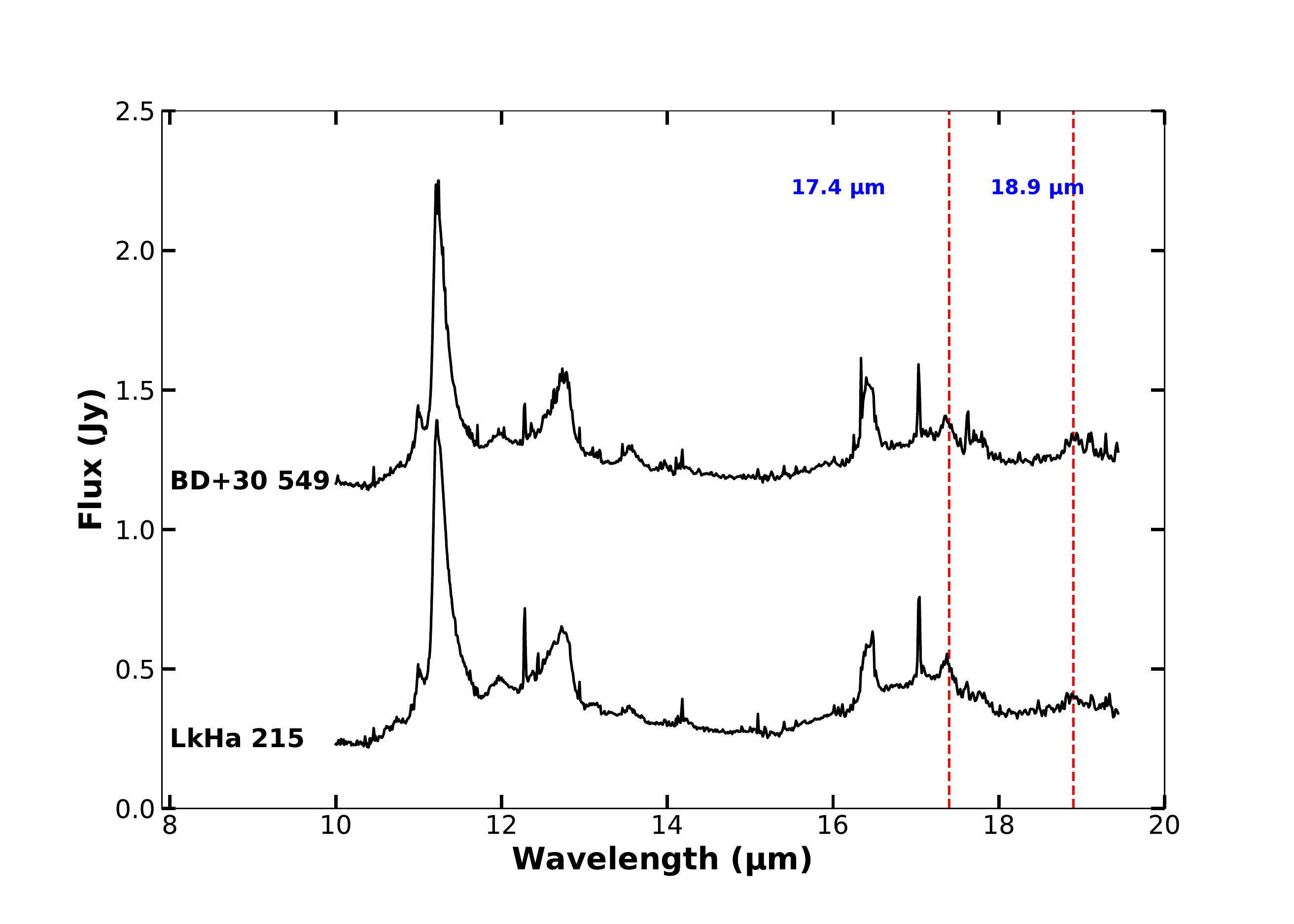}
    \caption{The figure shows the high-resolution Spitzer IRS spectra of two HAeBe stars with C\textsubscript{60} emission features. An offset of 1 Jy is provided in the flux values of BD+30 549 for better visualization of the spectra. The two vibrational modes of C\textsubscript{60} are shown as red lines.}
    \label{fig:high}
\end{figure*}

\subsection{Investigation of IRS spectra of HAeBe stars to search for $C_{60}$ features}

The IRS spectra of 126 HAeBe stars were searched to identify broad C\textsubscript{60} molecular emission. The 7.0 and 8.5 $\mu m$ vibrational modes are heavily contaminated by the strong PAH emission features seen in HAeBe stars. In six spectra, we observed the emission bands in the 16-20 $\mu m$ region called ``17 $\mu m$ plateau'' \citep{VanKerckhoven2000A&A...357.1013V,Shannon2015ApJ...811..153S}. The plateau is made up of 16.4 $\mu m$ PAH feature, H\textsubscript{2} at 17 $\mu m$ and C\textsubscript{60} at 17.4 $\mu m$ \citep{Iglesias2019MNRAS.489.1509I}. The C-C-C bending modes of the PAH feature are reported to be responsible for the 17 $\mu m$ plateau \citep{Allamandola1989ApJS...71..733A,Peeters2004ApJ...617L..65P} and the 17.4 $\mu m$ is a blend of PAH, and C\textsubscript{60} features \citep{Peeters2012ApJ...747...44P}. The 18.9 $\mu m$ emission is free from contamination and blending effects by other species. We prioritised the 18.9 $\mu m$ feature for the investigation.

We identified nine low-resolution and two high-resolution IRS spectra with 17.4 and 18.9 $\mu$m vibrational modes of C\textsubscript{60}. Amongst the sample, two stars have both high and low-resolution spectra. The C\textsubscript{60} features are visible in both of these spectra. \autoref{tab:full_table} shows the observation log of nine IRS spectra detected with C\textsubscript{60} features. The low-resolution IRS spectra of HAeBe stars identified with C\textsubscript{60} features are shown in \autoref{fig:1}. The fluxes are arbitrarily scaled for the purpose of display. The spectra used to identify C\textsubscript{60} features in all the 126 HAeBe stars are default recommendations (best extraction) by CASSIS. We individually compared the spectra generated from both extraction routines, i.e. optimal and tapered column, for all nine HAeBe stars identified with C\textsubscript{60}. All but one star have the best extraction as `tapered column' extraction, which appears to be suited for further analysis. The best extraction given by CASSIS for star CPM 25 is `optimal extraction'. But we find that tapered column extraction gives better spectra visually. Also, CPM 25 is associated with an RNe, which means the emission from the region has an extended nature.

 We visually inspected the full 5--20 $\mu m$ window in these nine spectra. C\textsubscript{60} vibrational modes at 7 and 8.5 $\mu m$ are reported to be identified in combination with 6--9 $\mu m$ plateau that are due to Hydrogenated Amorphous Carbons (HACs) \citep{Garc2010ApJ...724L..39G} or PAHs in the 5--15 $\mu m$ region. We identified broad PAH emission in the 5--15 $\mu m$ region and 17 $\mu m$ plateau in five spectra. Interestingly, the 6--9 $\mu m$ plateau due to HACs, which is commonly found in typical PNe spectra, was observed in two stars, namely HD 319896 and SAO220669. Additionally, we found only the C\textsubscript{60} features in the 15--20 $\mu m$ region of their spectra. In the case of HD 319896, the 7.0 $\mu m$ vibrational mode is visually detected, but the 8.5 $\mu m$ feature is not found.  
 
 Regarding the two stars, HD 46060 and MWC 593, the flux levels in the 5--15 $\mu m$ region of their spectra are considerably less than the 15--20 $\mu m$ region. Also, none of the broad features is identified in the 5--15 $\mu m$ region. HD 46060 shows 17 $\mu m$ plateau along with 18.9 $\mu m$ C\textsubscript{60} feature. MWC 593 shows only the C\textsubscript{60} features at 17.4 and 18.9 $\mu$m.

 The detection of C\textsubscript{60} in high-resolution spectra of LkHa 215 and BD+30 549 confirms the detection on the low-resolution spectra. The high-resolution IRS spectra of two HAeBe stars are shown in \autoref{fig:high}. Until now, the C\textsubscript{60} features are primarily identified near evolved stars. And very few star-forming regions are reported to have C\textsubscript{60} features. Thus, individual attention is given to the information available in the literature for each HAeBe star to confirm their YSO nature. Detailed information about objects of interest is listed in the Appendix B. We evaluated the relevant literature and confirm the HAeBe nature of the nine stars used in this study. The stellar parameters of the HAeBe stars, such as distance, age, and mass, are listed in \autoref{tab:full_table}. Also, their spectral features are listed in \autoref{tab:full_table}. The five stars which showed intense PAH emission features in their IRS spectra are associated with RNe \citep{Magakian2003A&A...399..141M}.

\subsection{MIR analysis using IRS spectra}

The comparison of IRS spectra of HAeBe stars with the known C\textsubscript{60} emitting sources and the estimation of continuum spectral indices are described in this section. 
\begin{figure*}
    \includegraphics[width=0.66\columnwidth]{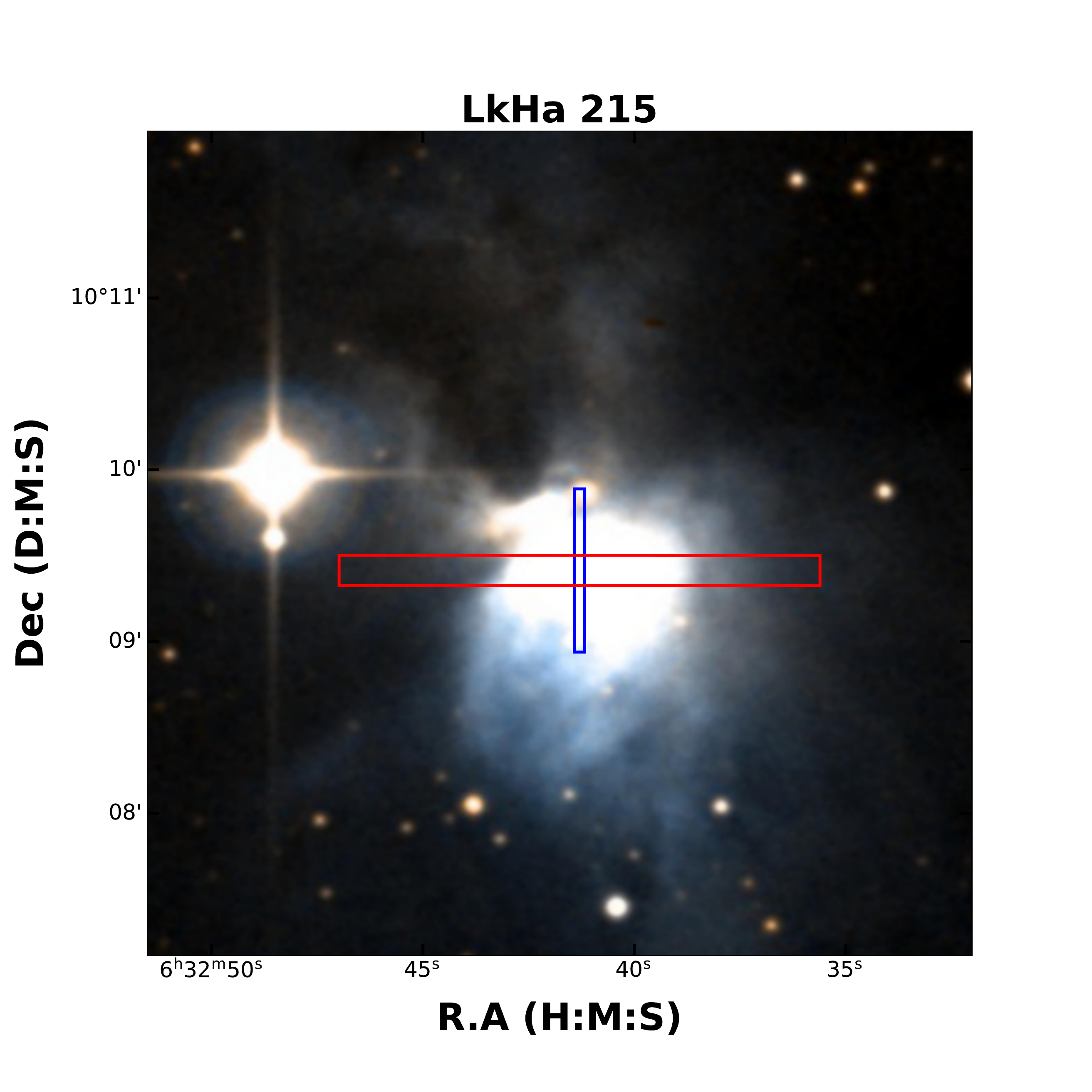}
    \includegraphics[width=0.66\columnwidth]{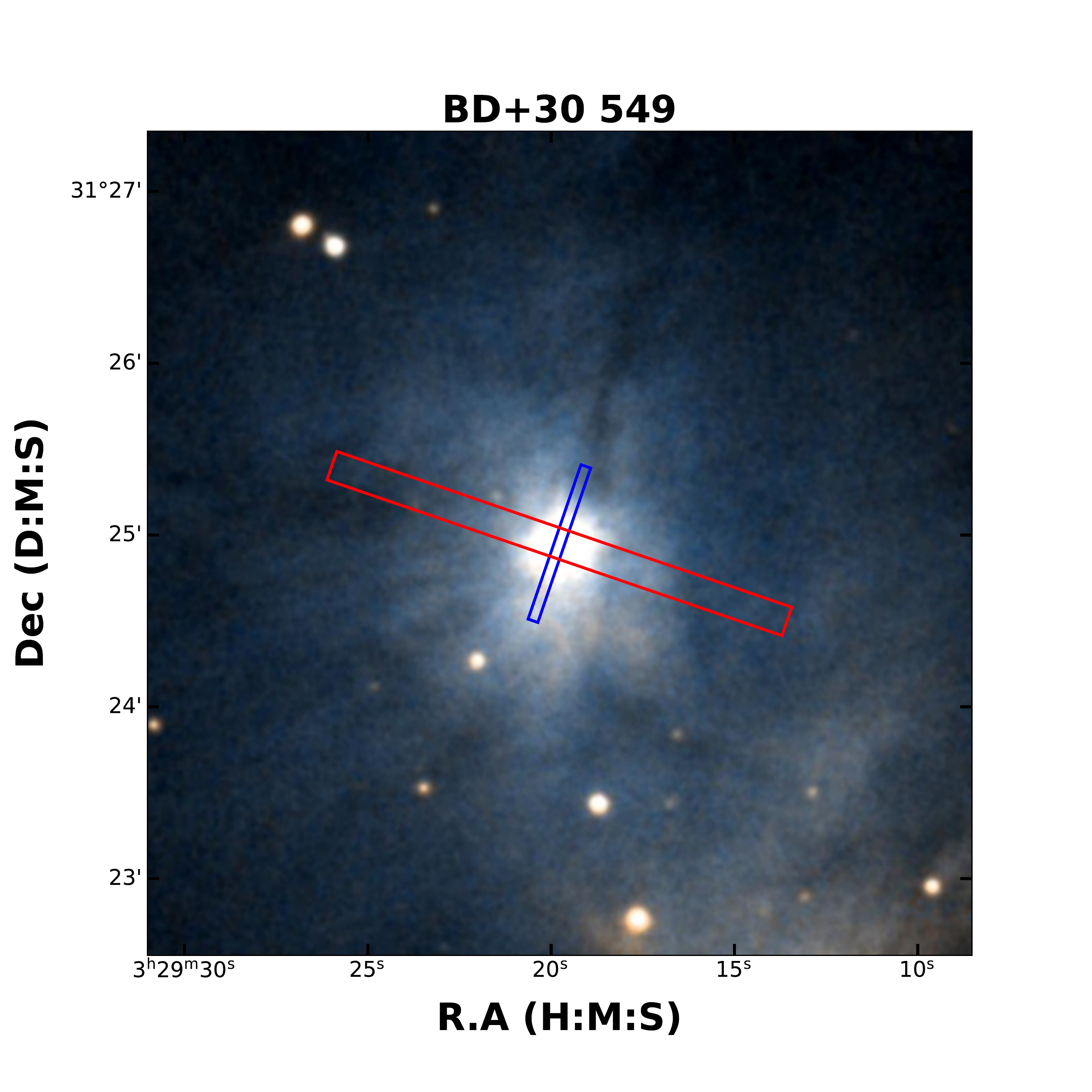}
    \includegraphics[width=0.66\columnwidth]{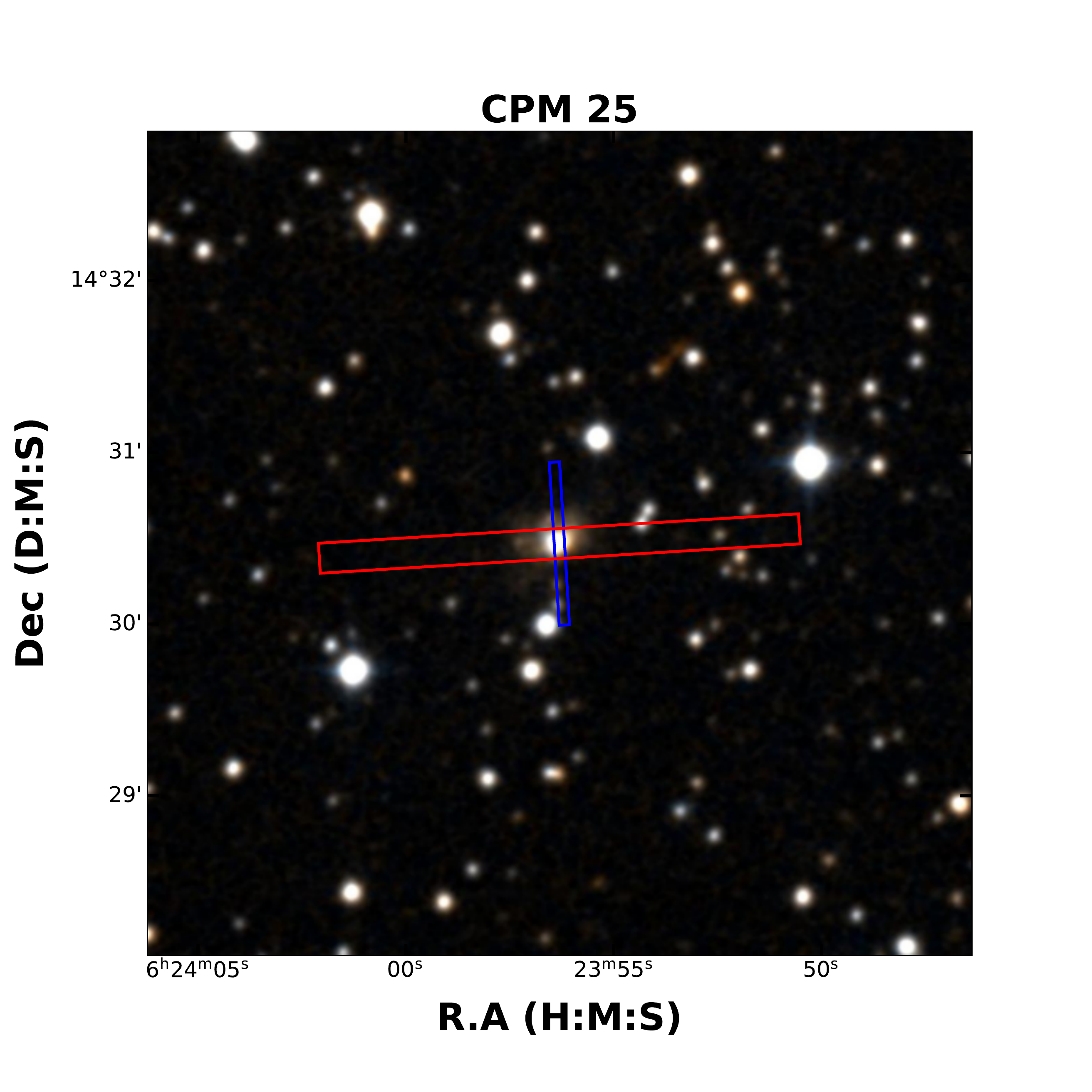}
    \includegraphics[width=0.66\columnwidth]{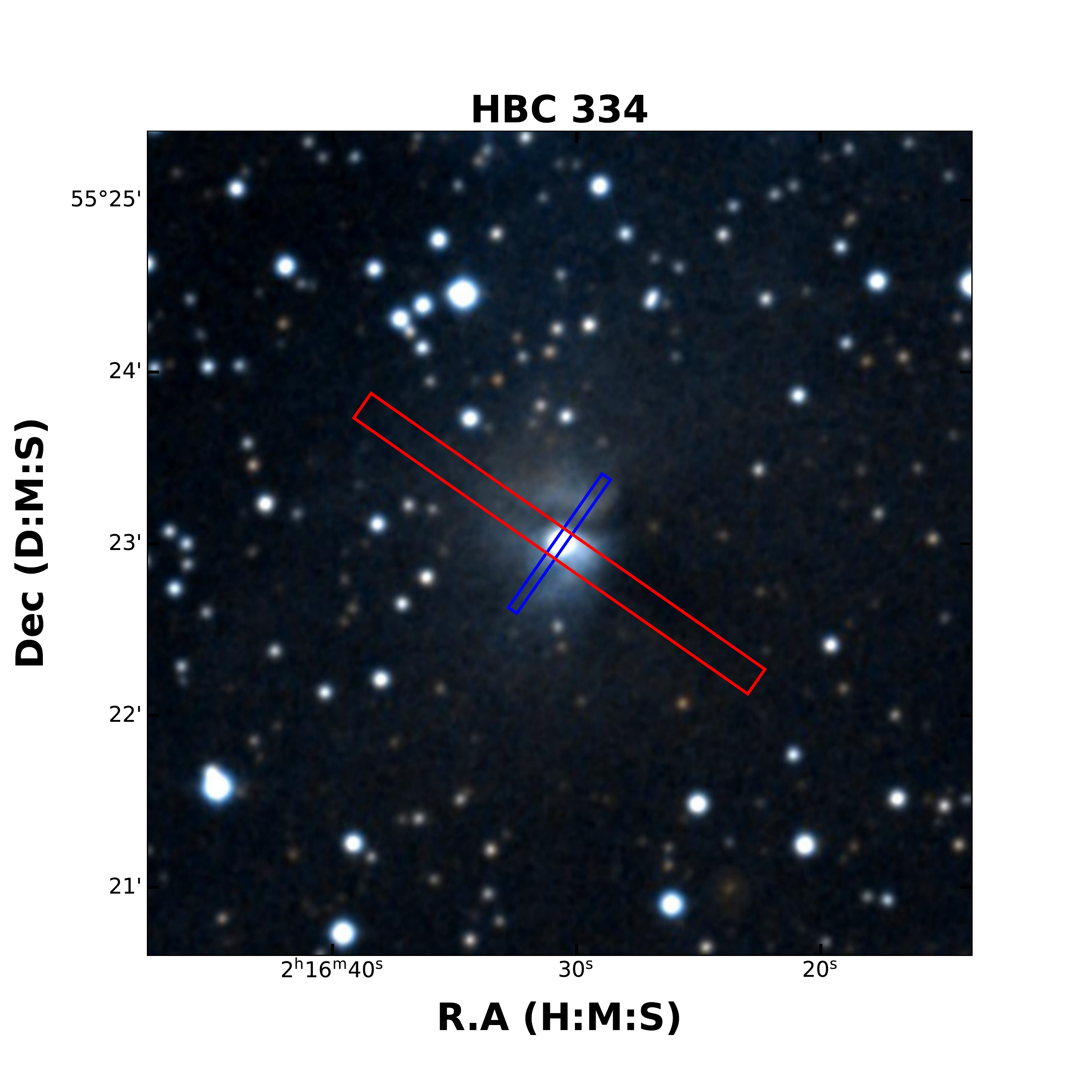}
    \includegraphics[width=0.66\columnwidth]{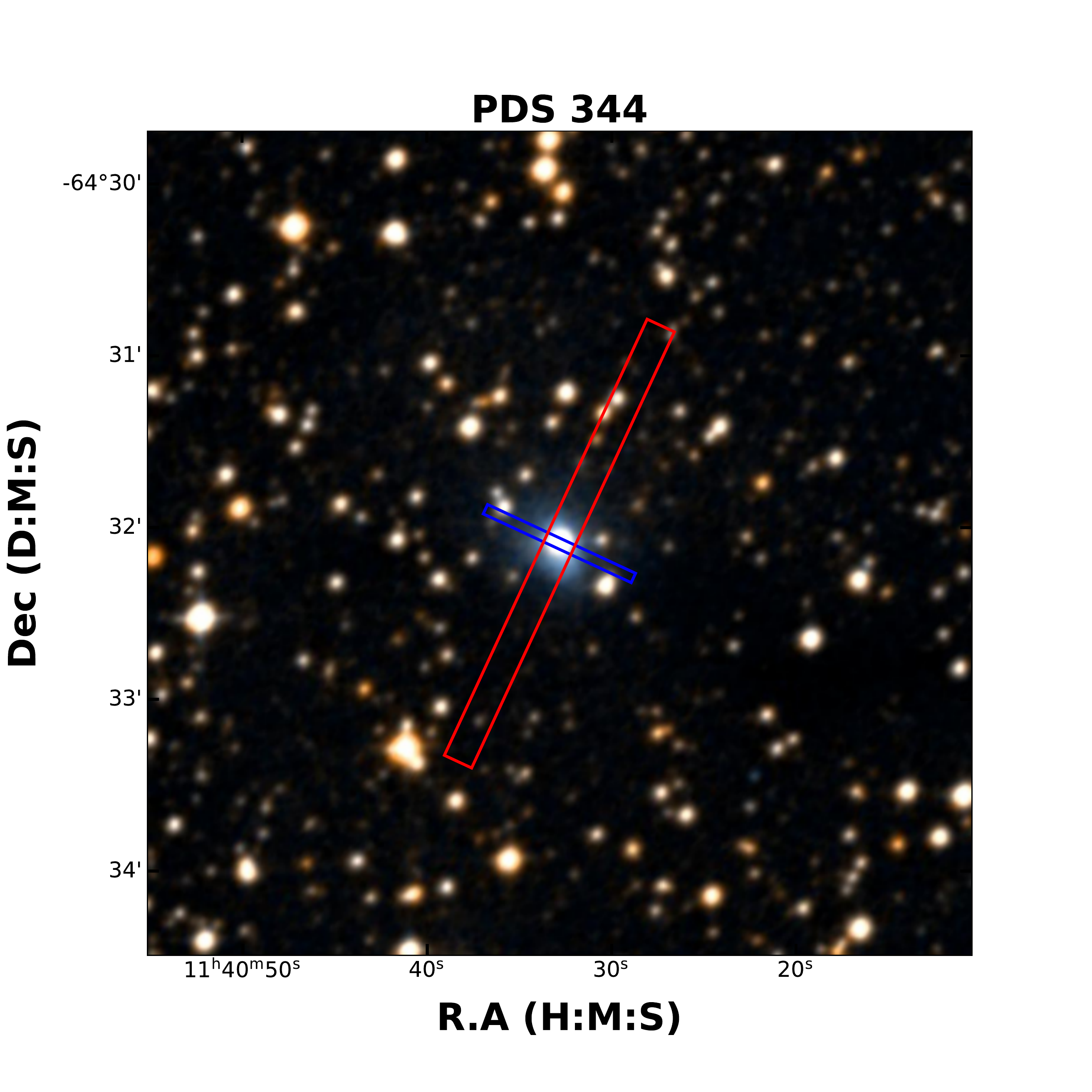}
    \includegraphics[width=0.66\columnwidth]{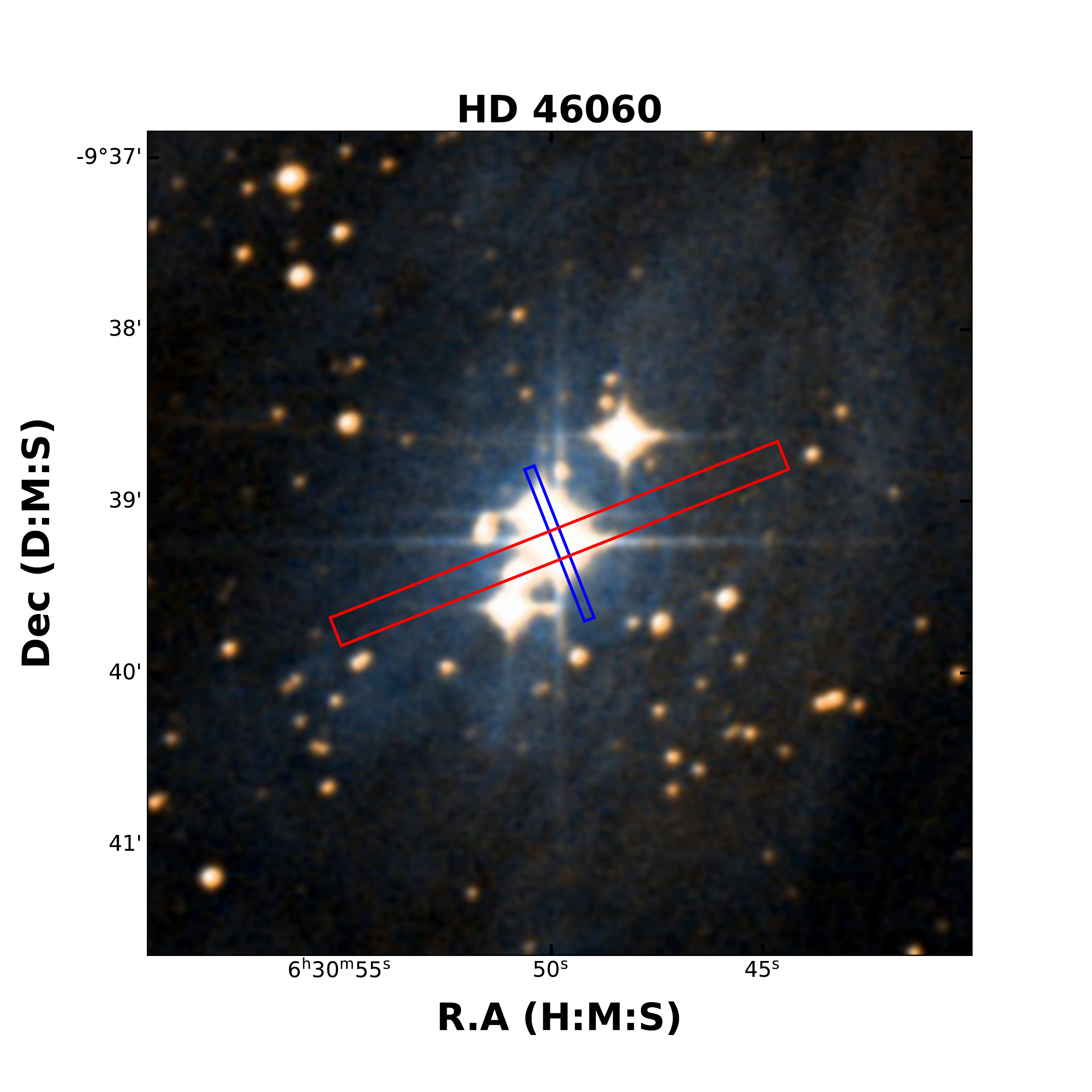}
    \includegraphics[width=0.66\columnwidth]{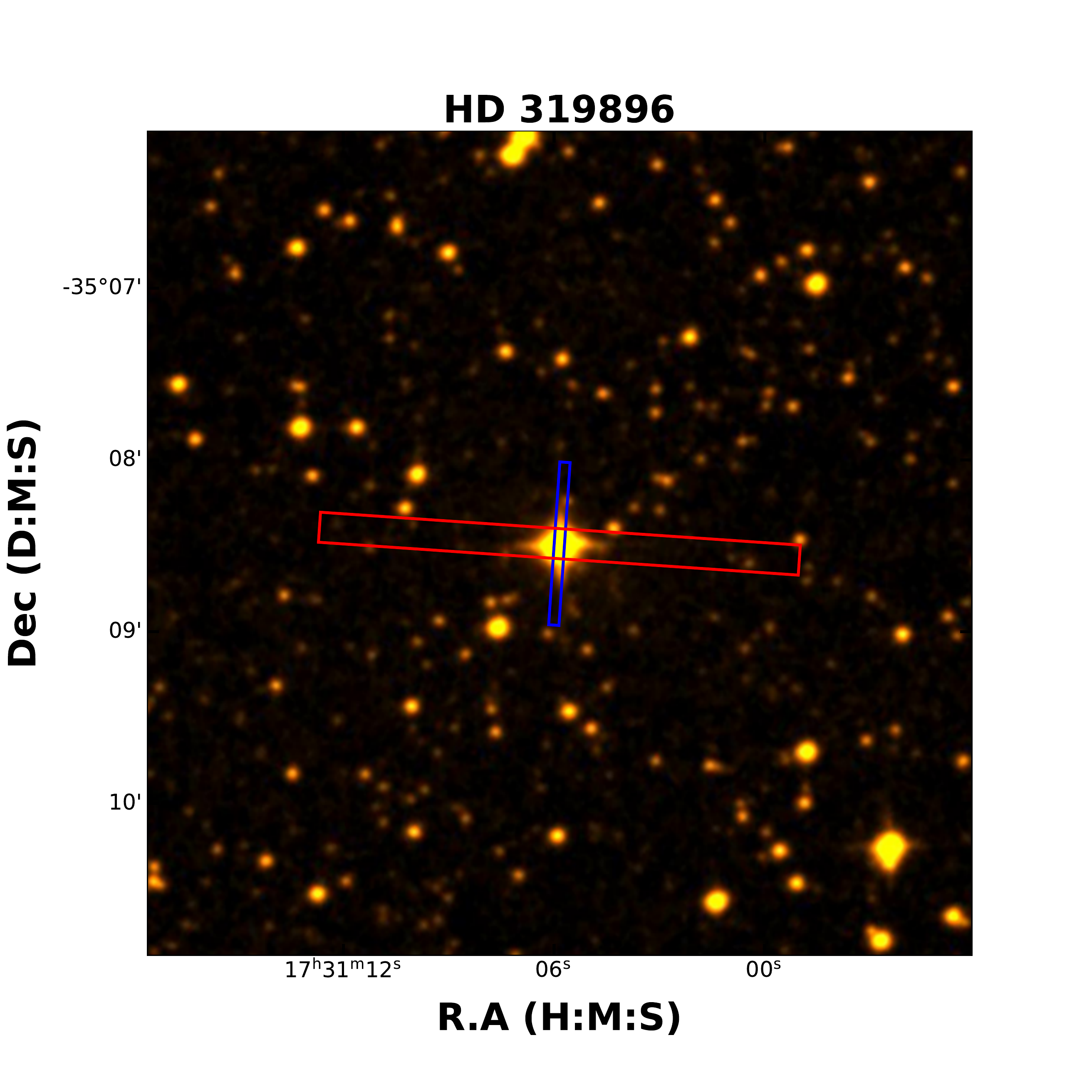}
    \includegraphics[width=0.66\columnwidth]{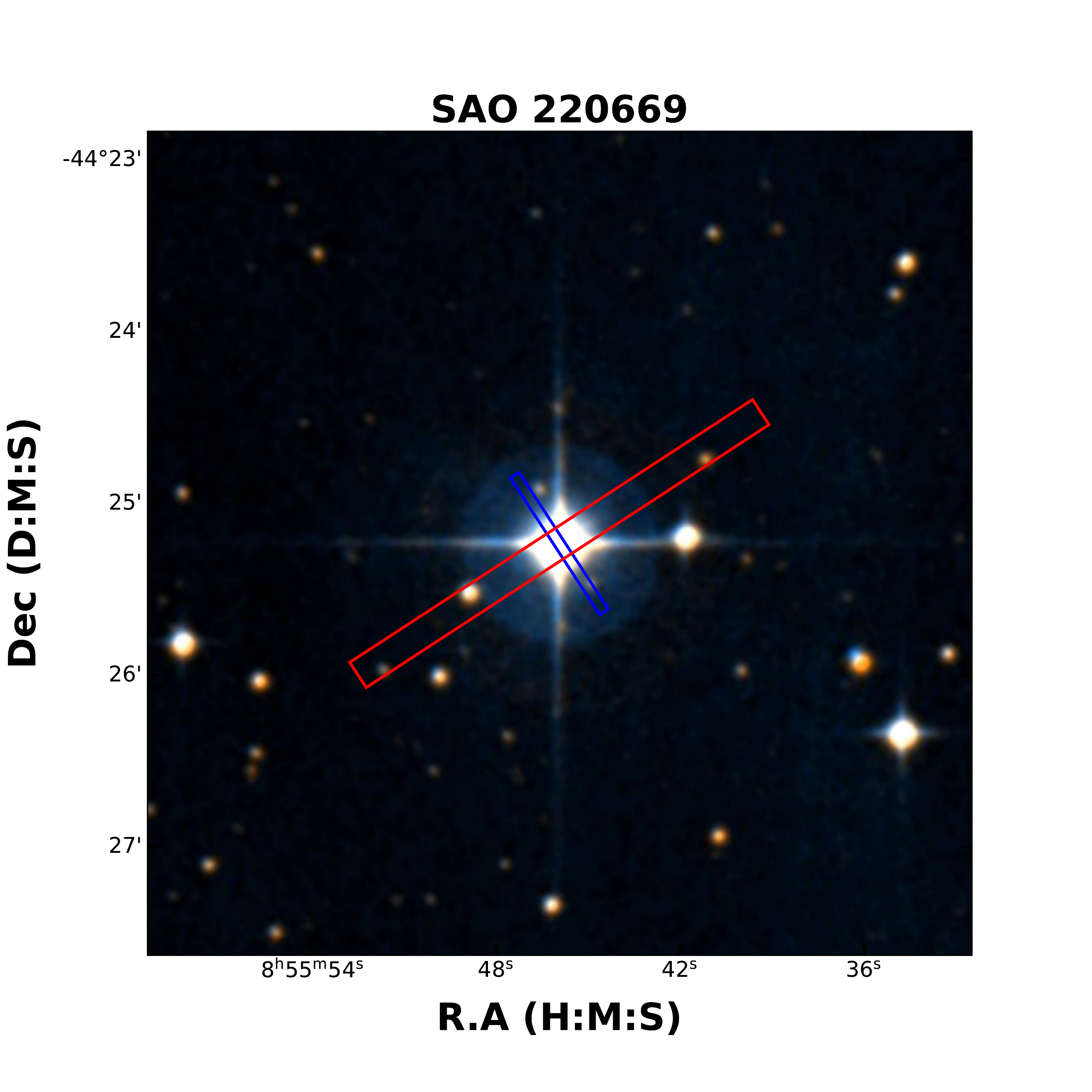}
    \includegraphics[width=0.66\columnwidth]{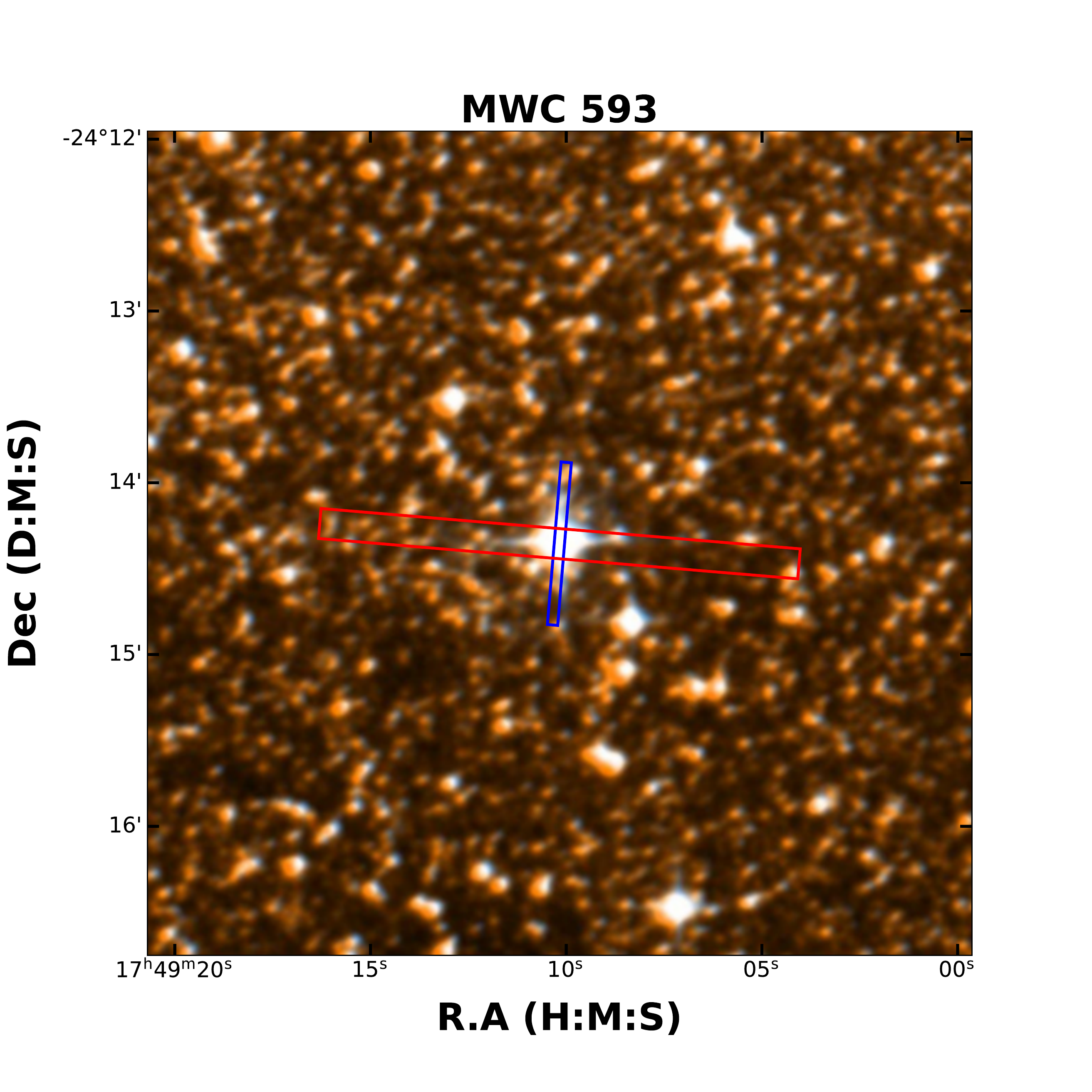}
    \caption{The DSS2/color images of nine HAeBe stars with the IRS slits are shown in the figure. First, six sources are associated with RNe. The three sources in the bottom panel were not found to have reflection nebulosity. The red and blue rectangles are SL and LL slit orientations respectively.}
    \label{fig:imagesl}
\end{figure*}

\subsubsection{Comparison of IRS spectra of HAeBe stars with the known C\textsubscript{60} emission sources}

We carefully examined the continuum subtracted spectra of nine HAeBe stars and found that five of them exhibit similar spectral features, characterized by a 17 $\mu$m plateau and intense PAH emission features in the 5--15 $\mu m$ region. Another star, HD 46060, also shows a 17 $\mu$m plateau, but with insufficient flux in the 5-15 $\mu m$ region. Interestingly, these six stars are associated with RNe \citep{Magakian2003A&A...399..141M}. The DSS color images of the HAeBe stars observed with the IRS slits are shown in \autoref{fig:imagesl} (first six), revealing the nebulosity around them. We compared these six HAeBe stars to known sources of C\textsubscript{60} emission features associated with RNe.

\begin{figure*}
    \centering
    \includegraphics[width=2\columnwidth]{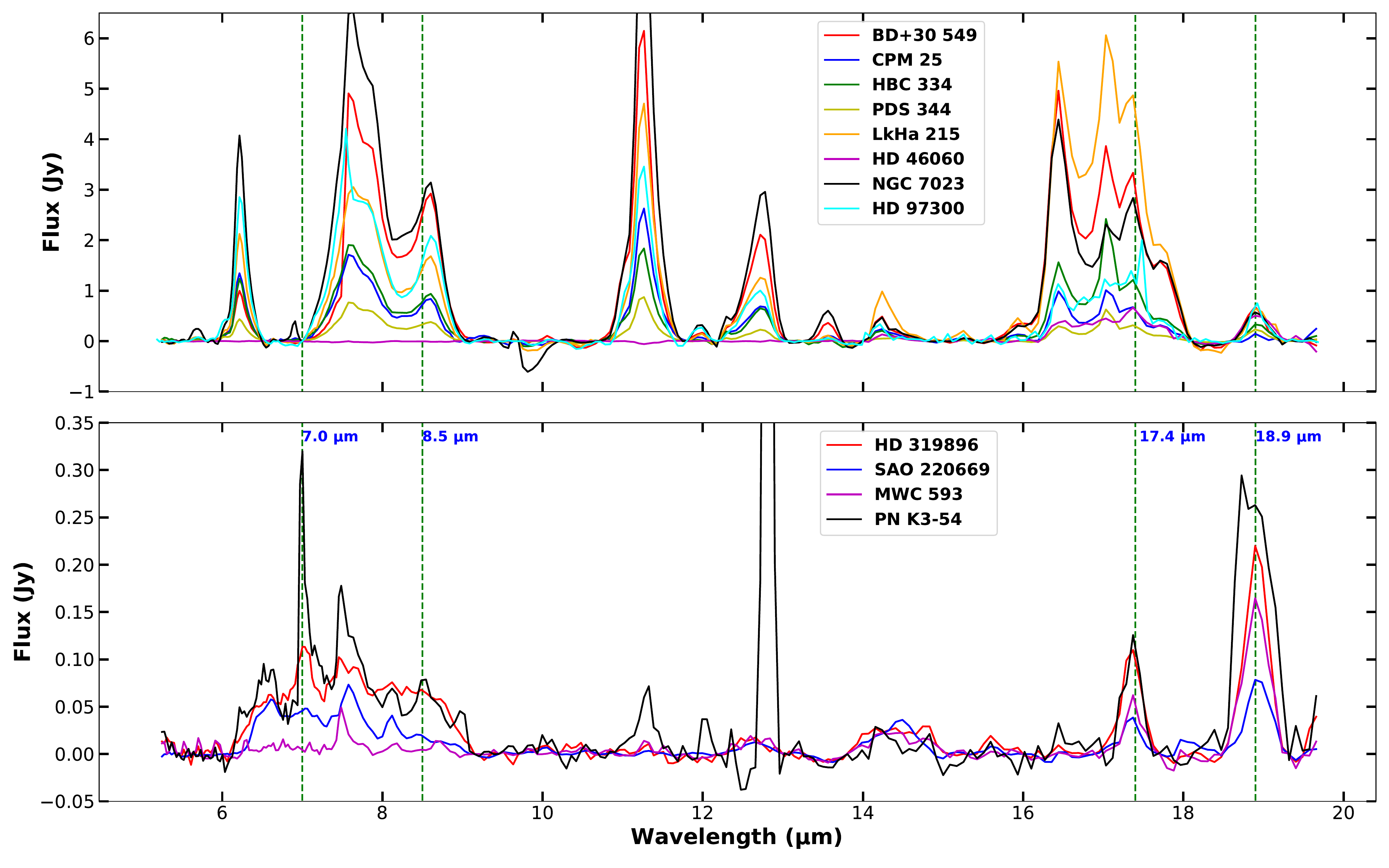}

    \caption{The figure shows the IRS spectra classified into two groups. The ``RNe-like" spectra are shown in the top panel and the ``PNe-like" spectra in the bottom panel. The IRS spectra of RNe NGC 7023, HD 97300 (top panel) and PNe K3-54 (bottom panel) are also compared. }
    \label{fig:fullrnelike}
\end{figure*}

We obtained the IRS spectrum of HD 97300, the first HAeBe star known to exhibit C\textsubscript{60} features, from \cite{Manoj2011ApJS..193...11M}, and used the low-resolution IRS spectrum of RNe NGC 7023 (AORkey - 3827712) from CASSIS. Notably, NGC 7023 is illuminated by the HAeBe star HD 200775 \citep{Saha2020}. Our sample of stars exhibiting C\textsubscript{60} emission features were compared with the spectra of HD 97300 and RNe NGC 7023, as shown in \autoref{fig:fullrnelike}. The spectra exhibit strong PAH features at 6.2, 7.7, 8.6, 11.2, and 12.6 $\mu m$ in the 5--15 $\mu m$ region, and the 15--20 $\mu m$ region shows a 17 $\mu m$ plateau. This plateau was previously identified in the RNe NGC 2023 and is attributed to large PAHs \citep{Peeters2012ApJ...747...44P}. The similarity between the spectra of the six HAeBe stars and HD 97300 and those of RNe NGC 7023, as well as their association with RNe, confirm that these objects have an "RNe-like" spectra and that the C\textsubscript{60} features originate from the nebulosity associated with the HAeBe stars.

Among the sample of nine HAeBe stars, three stars - HD 319896, SAO 220669, and MWC 593 - show only C\textsubscript{60} features in the 15--20 $\mu m$ region and no intense PAH features as seen in the "RNe-like" spectra of the other six stars. HD 319896 and SAO 220669 exhibit the 6--9 $\mu m$ plateau, attributed to HACs, which is also seen in the PNe K3-53 spectrum. However, due to low flux, MWC 593 shows no features in the 5-15 $\mu m$ region. These three HAeBe stars are not associated with RNe. Comparing their spectra with the known C\textsubscript{60} detected PNe K3-54 shows that they have similar IRS spectral features such as the 6--9 $\mu m$ plateau, 17.4 and 18.9 C\textsubscript{60} features. The C\textsubscript{60} feature at 7 $\mu m$ and 8.5 $\mu m$ is seen in the PNe K3-54 spectra. HD 319896 is the only HAeBe star detected in our sample with the 7 $\mu m$ micron feature. However, the presence of intense forbidden [NeII] emission at 12.81 $\mu m$ distinguishes the PNe spectra from the spectra of HD 319896, and SAO 220669, which are labelled as "PNe-like" due to the common C\textsubscript{60} features initially observed in PNe spectra. The DSS color images of the three HAeBe stars with the IRS slits are shown in \autoref{fig:imagesl} (bottom three).

\subsubsection{Continuum Spectral Indices}
\label{sec:index}

In this section, we have analyzed the MIR spectra of HAeBe stars to investigate the distinctiveness of their spectra based on the presence and absence of C\textsubscript{60} and nebulosity. We have used two continuum spectral indices, $n_{2-25}$ and $n_{5-12}$, to evaluate this distinctiveness. The former is used to classify the stars into Class I, Flat spectrum (FS), Class II, and Class III sources \citep{LAda1987, Wilking1989, Green1994,Manoj2011ApJS..193...11M}, while the latter is an extinction-free index that characterizes the slope of the SED between 5-12 $\mu m$ \citep{McClure2010ApJS..188...75M, Manoj2011ApJS..193...11M}. We selected the 55 HAeBe stars low-resolution IRS spectra with 5-36 $\mu m$ window used in the analysis. 

The first index, $n_{2-25}$, was computed using the 2MASS K\textsubscript{s} magnitude and the 25 $\mu m$ flux obtained from the IRS spectra. We acknowledge that the empirical SED classification using $n_{2-25}$ index may be affected by viewing geometry and line-of-sight extinction \citep{Robitaille2007ApJS..169..328R,Crapsi2008A&A...486..245C, McClure2010ApJS..188...75M}. The second index, $n_{5-12}$, was also estimated, known as an extinction-free index. This index was derived from the slope of the SED estimated between wavelengths of 5-12 $\mu m$, and it is not affected by extinction.

\begin{figure}
    \centering
    \includegraphics[width=1.1\columnwidth]{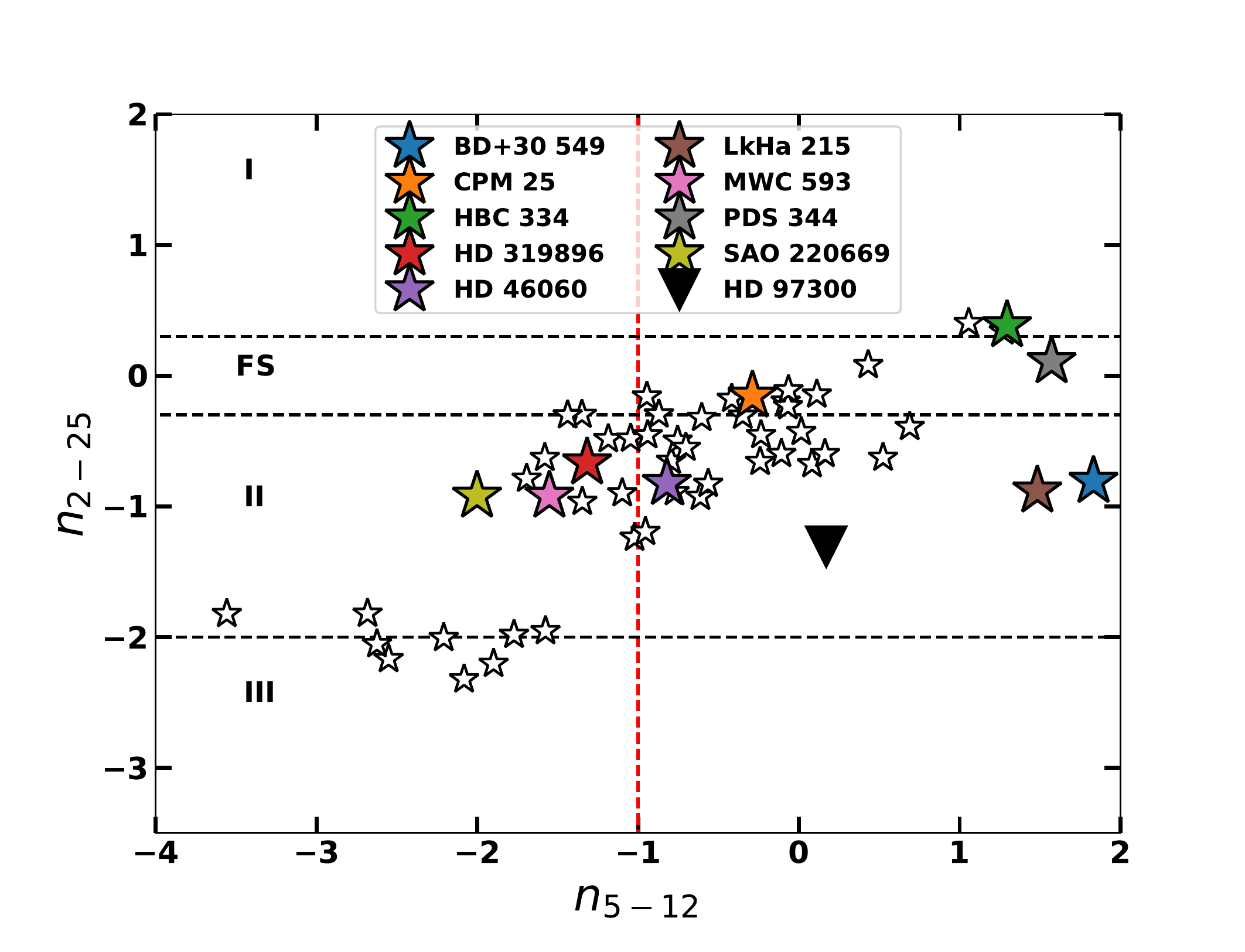}
 
    \caption{The figure shows the continuum spectral index comparison of the sample of IRS spectra used in this study. The spectral index of HAeBe stars with C\textsubscript{60} features are illustrated in various colors. Also, HD 97300, the previously known HAeBe star with C\textsubscript{60} features, is shown in the figure. The 46 HAeBe stars with IRS spectra used in the study are shown in open star symbols. The red line denotes $n_{5-12}$ = -1, which divides HAeBe stars associated with nebulosity. Horizontal black dashed lines shows the Class I, FS, Class II, and Class III classification of YSOs using the index, $n_{2-25}$.}
    \label{fig:full2}
\end{figure}

\autoref{fig:full2} illustrates the distribution of both spectral indices, $n_{2-25}$ and $n_{5-12}$, for the HAeBe stars, along with their classification according to $n_{2-25}$ as Class I, FS, Class II, and Class III sources. The remaining 46 HAeBe stars with 5-38 $\mu m$ IRS spectra are also shown in the figure as reference stars. For the purpose of investigating the dominant source of MIR emission, \cite{McClure2010ApJS..188...75M} proposed a classification scheme using $n_{5-12}$ index for T Tauri stars, which helps to distinguish between photospheric, disk, and envelope-dominated MIR emission. In this study, we apply this index to check for differences between sources with "RNe-like" and "PNe-like" spectra.

The analysis of \autoref{fig:full2} reveals that all HAeBe stars with C\textsubscript{60} are Class II and above. We also include HD 97300 in the figure for reference. Interestingly, "RNe-like" sources have a spectral index $n_{5-12}$ $>$ -1. It is worth noting that the IRS spectra of RNe NGC 2245 exhibit a spectral index of $n_{5-12}$ as 0.2. We investigated the possibility of HAeBe stars associated with RNe having a higher $n_{5-12}$ index. Among the 55 HAeBe stars used in this study, we identified 13 stars with an index of $n_{5-12}$ $>$ -0.5 and an effective temperature greater than 10,000 K. Among these 13 stars, 10 show an association of nebulosity. However, HD 319896, MWC 593, and SAO 220669 have $n_{5-12}$ $<$ -1 and are not associated with nebulosity. This observation implies that the MIR emission in these spectra is not originating from an extended nebulous region but a much closer region such as circumstellar disks.

\section{DISCUSSION}

To date, C\textsubscript{60} emission features have primarily been identified in planetary nebulae (PNe) objects, with only a few instances of identification in embedded young stellar objects (YSOs), one Herbig Ae/Be star (HD 97300), and a couple of RNes. However, a comprehensive search for C\textsubscript{60} features in YSOs was absent in the literature until now. In this study, we present a MIR spectral catalog for 126 HAeBe stars and identify nine HAeBe stars with C\textsubscript{60} emission features. We discuss the properties of the identified sample in detail, shedding new light on the prevalence and characteristics of C\textsubscript{60} in YSOs.

\subsection{Detection fraction of C\textsubscript{60} in HAeBe stars}

In this study, we analyzed a sample of 126 Spitzer IRS spectra of Herbig Ae/Be (HAeBe) stars, providing the first systematic search for C\textsubscript{60} emission features in this population. Our analysis revealed C\textsubscript{60} emission features in 9 of the IRS spectra, corresponding to a detection fraction of approximately 7\%. However, as our sample represents only 50\% of the total population of HAeBe stars, and with the number of HAeBe stars increasing from recent survey programs, such as \citet{Zhang2022ApJS..259...38Z}, \citet{Shridharan2021RAA....21..288S}, and \citet{Vioque2020A&A...638A..21V}, the detection fraction of C\textsubscript{60} in HAeBe stars may increase with future observations.

Furthermore, we found that six HAeBe stars identified with C\textsubscript{60} emission are associated with reflection nebulae (RNe), with a detection fraction of 30\% among the 20 HAeBe stars in our sample that are associated with RNe. Considering HD 97300 and HD 200775, already known to be associated with RNe, the detection fraction may increase with increased observations near PDRs. This finding supports the notion that C\textsubscript{60} is more common in RNe and photon-dominated regions (PDRs) \citep{Castellanos2014ApJ...794...83C}, and the formation of C\textsubscript{60} in the ISM is as dominant as in the expanding envelopes of PNe. Future observations, particularly with the JWST, may lead to an increased detection fraction of C\textsubscript{60} emission in HAeBe stars and provide further insight into the prevalence and distribution of fullerenes in different astrophysical environments.

\subsection{Herbig Be stars: Fullerene factories for ISM}
The spectral type distribution of the nine HAeBe stars that shows C\textsubscript{60} emission provides some interesting insights. While HAeBe stars have a wide range of spectral types (B0-F5; \citealp{WATERS1998}), the fact that the identified stars in this study have spectral types B9 or earlier indicates that the UV radiation from the central star is crucial in the excitation of C\textsubscript{60}. This finding is consistent with previously known HAeBe stars with C\textsubscript{60}, such as HD 200775 and HD 97300, which have spectral types B2 and B9, respectively. It is noteworthy from the largest IRS spectral catalog, only B-type intermediate-mass YSOs host C\textsubscript{60} in their circumstellar region. At the same time, no detection of C\textsubscript{60} has been reported near A-type HAeBe stars or T Tauri stars.

C\textsubscript{60} can form through two main scenarios: photo-chemical processing of large PAH molecules \citep{Bern2012PNAS..109..401B} and HACs \cite{Garc2010ApJ...724L..39G}. In the first scenario, when large PAH molecules are subjected to UV photons, they undergo dehydrogenation to form Graphene, which subsequently loses Carbon to form C\textsubscript{60} \citep{Bern2012PNAS..109..401B,Zhen2014ApJ...797L..30Z}. This pathway has been observed in RNe such as NGC 7023, where the calculated Carbon and Hydrogen loss rates are consistent with the age of the nebulae \citep{Bern2012PNAS..109..401B}. The spectra of six HAeBe stars with RNe-like characteristics show similarities with NGC 7023, indicating the presence of abundant PAH molecules in their environment, as evidenced by strong PAH emission features in their IRS spectra. The 17 $\mu m$ plateau is attributed to larger PAH molecules \citep{Peeters2012ApJ...747...44P}. PAHs with 60 or more Carbon atoms are essential for this pathway to be viable. Laboratory experiments have shown that the top-down formation of C\textsubscript{60} from PAHs is possible \citep{Zhen2014ApJ...797L..30Z}. Recent modelling studies by \cite{Murga2022MNRAS.517.3732M} also support this formation pathway.

The aforementioned scenario has been proposed to occur in the inner regions of circumstellar disks around young stars undergoing magnetospheric accretion, where intense UV radiation can be generated \citep{Bern2012PNAS..109..401B}. However, the HAeBe stars identified in this study all have spectral types of B or earlier, indicating that the top-down formation of C\textsubscript{60} from PAHs dominates in less dense circumstellar environments, such as the circumstellar medium (RNe) illuminated by Herbig Be stars, compared to the circumstellar disks around Herbig Ae stars. The study shows B type pre-main sequence stars hosts perfect environment for the formation and population of C\textsubscript{60} in the ISM.

\subsection{Possible detection of circumstellar C\textsubscript{60}}

HD 319896 and SAO 220669 are two stars with spectra similar to PNe, displaying a plateau between 6-9 $\mu m$ attributed to HACs. \cite{Garc2010ApJ...724L..39G} proposed that HACs can be photochemically processed to produce the C\textsubscript{60}. The processing of HACs can also be a viable pathway for C\textsubscript{60} formation. MWC 593 also exhibits a similar spectrum, but lacks the 6-9 $\mu m$ plateau. Also, none of these stars have been associated with nebulosity, nor do they exhibit the characteristics of evolved objects. A similar conclusion is drawn from the $n_{5-12}$ index analysis (Sect 3.3.2.). Therefore, it is likely that the emission features of  C\textsubscript{60} originate from a much closer structure, such as a circumstellar disk.

We retrieved the optical spectra of HD 319896 from the ESO archive, which was observed using the HARPS spectrograph \citep{Pepe2000SPIE.4008..582P, Mayor2003Msngr.114...20M}. The H$\alpha$ emission profile shows a double peak structure, indicating the presence of a circumstellar disk. SAO 220669 shows no emission lines in its spectra taken from VLT/X-shooter spectrograph \citep{Vernet2011A&A...536A.105V} but has IR excess, which is suggestive of the presence of a circumstellar disk. 

However, the spatial extent of MIR emission calculated by CASSIS for HD 319896, SAO 220669, and MWC 593 are 11.9", 6.7", and 4.9", respectively. We checked the spatial profiles of the three sources. HD 319896 is contaminated by a second source in the slit. But in the LL slit, which has 17.4 and 18.9 $\mu m$ emission, the contamination is not seen and the super-sampled PSF is fitting the source emission. The higher spatial extent for HD 319896 might be due to the presence of the secondary source. The other two sources are not affected by any contaminating source. The nature of the emission is slightly extended as tapered column extraction suggests. There could be diffused emission regions close to the star if not "MIR bright" circumstellar disk around these stars. More dedicated and deep Integral Field Spectroscopic (IFU) observations from {\it JWST} could confirm the origin of the emission.

The analysis of six HAeBe stars reaffirms that the C\textsubscript{60} features are seen near YSOs and are common near RNes. They present a unique opportunity to solve the mystery of formation and excitation of C\textsubscript{60} in environments other than that of PNes. The future studies of these HAeBe stars and their environment, especially the spectral and imaging studies using {\it JWST} can provide more insights in this direction.

\section{Conclusions}

In this study, we report the detection of C\textsubscript{60} spectral features in the IRS spectrum of 9 HAeBe stars. The main conclusions of this work are summarized as follows:

We created a catalog containing the IRS spectra of 126 HAeBe stars using the CASSIS database $-$ the largest MIR spectral catalog of HAeBe stars to date. We searched for the vibrational modes of C\textsubscript{60} in these spectra and identified nine HAeBe stars with the C\textsubscript{60} molecules in their circumstellar medium.

Our MIR analysis revealed that six HAeBe spectra are associated with reflection nebulosity (RN). These spectra exhibited similar spectral characteristics to previously studied RNe NGC 7023. The C\textsubscript{60} emission features in these stars originate from the nebulous circumstellar region. Three stars without any association of RN showed similarities with the features seen in planetary nebulae (PNe) MIR spectra. The origin of C\textsubscript{60} emission features in these stars may be from nearby diffused emission regions or the circumstellar disk.

The detection fraction of C\textsubscript{60} in the sample of HAeBe stars is around 7\%. However, the detection fraction increases to 30\% when considering only the HAeBe stars with nebulosity. In this catalog, we found that only B-type HAeBe stars host C\textsubscript{60} in their circumstellar region, while no detection of C\textsubscript{60} has been reported near A-type HAeBe stars.

The catalog and the detection of C\textsubscript{60} in HAeBe stars can act as a reference for future JWST proposals, which will help advancing our understanding of the physical processes and chemical composition of the circumstellar environments of HAeBe stars.

\section*{Acknowledgements}

We would like to thank the anonymous referee for providing helpful comments and suggestions that improved the paper. The authors are grateful to Dr. Vianney Lebouteiller for discussions on the CASSIS archive. RA acknowledges Prof. Annapurni Subramaniam for the financial support from SERB POWER fellowship grant SPF/2020/000009. BM and BS acknowledges the support of the SERB, a statutory body of the Department of Science \& Technology (DST), Government of India, for funding our research under grant number CRG/2019/005380. 

%%%%%%%%%%%%%%%%%%%%%%%%%%%%%%%%%%%%%%%%%%%%%%%%%%
\section*{Data Availability}

The information used in this article was obtained from the Combined Atlas of Sources with Spitzer IRS Spectra (CASSIS) and is available at https://cassis.sirtf.com/. The results produced by this study will be made available to interested parties upon request to the corresponding author.

%%%%%%%%%%%%%%%%%%%% REFERENCES %%%%%%%%%%%%%%%%%%

% The best way to enter references is to use BibTeX:

\bibliographystyle{mnras}
\bibliography{example} % if your bibtex file is called example.bib

% Alternatively you could enter them by hand, like this:
% This method is tedious and prone to error if you have lots of references
%\begin{thebibliography}{99}
%\bibitem[\protect\citeauthoryear{Author}{2012}]{Author2012}
%Author A.~N., 2013, Journal of Improbable Astronomy, 1, 1
%\bibitem[\protect\citeauthoryear{Others}{2013}]{Others2013}
%Others S., 2012, Journal of Interesting Stuff, 17, 198
%\end{thebibliography}

%%%%%%%%%%%%%%%%%%%%%%%%%%%%%%%%%%%%%%%%%%%%%%%%%%

%%%%%%%%%%%%%%%%% APPENDICES %%%%%%%%%%%%%%%%%%%%%

\appendix

\section{MIR catalog of HAeBe stars}

\onecolumn

\begin{longtable}{ccccccccc}
\caption{The table contains details of 126 HAeBe stars with Spitzer IRS spectra. The AOR\_LR and AOR\_HR are the Spitzer AOR key of low and high resolution observations respectively. Distance, T\textsubscript{eff}, LogL, and A\textsubscript{V} values are taken from Vioque et al. (2018).}
\label{tab:my-table}\\
\hline
Name & AOR\_LR & AOR\_HR & \begin{tabular}[c]{@{}c@{}}R.A\\ (Deg)\end{tabular} & \begin{tabular}[c]{@{}c@{}}Dec\\ (Deg)\end{tabular} & \begin{tabular}[c]{@{}c@{}}Distance\\ (pc)\end{tabular} & \begin{tabular}[c]{@{}c@{}}T\textsubscript{eff}\\ (K)\end{tabular} & LogL & \begin{tabular}[c]{@{}c@{}}A\textsubscript{V}\\ (mag)\end{tabular} \\ \hline
\endfirsthead
\multicolumn{9}{c}%
{{\bfseries Table \thetable\ continued from previous page}} \\
\hline
Name & AOR\_LR & AOR\_HR & \begin{tabular}[c]{@{}c@{}}R.A\\ (Deg)\end{tabular} & \begin{tabular}[c]{@{}c@{}}Dec\\ (Deg)\end{tabular} & \begin{tabular}[c]{@{}c@{}}Dist\\ (pc)\end{tabular} & \begin{tabular}[c]{@{}c@{}}Teff\\ (k)\end{tabular} & LogL & \begin{tabular}[c]{@{}c@{}}Av\\ (mag)\end{tabular} \\ \hline
\endhead
\hline
\endfoot
\endlastfoot
AK Sco & 12700160 &  & 253.686667 & -36.888611 & 140.6 & 6250 & 0.62 & 0 \\
AS 310 &  & 25733376 & 278.338333 & -4.968333 & 2108.4 & 24500 & 4.17 & 4.129 \\
AS 470 & 12683008 &  & 324.059167 & 57.358611 & 4039.6 & 8150 & 3.01 & 2.272 \\
BD+30 549 & 14121472 & 14121472 & 52.3325 & 31.415833 & 295.4 & 11500 & 1.54 & 1.727 \\
BF Ori & 18835968 & 5638144 & 84.305417 & -6.583611 & 388.8 & 8970 & 1.29 & 0.33 \\
BH Cep & 21886720 &  & 330.42875 & 69.743333 & 335.1 & 6630 & 0.76 & 0.831 \\
BO Cep & 21886976 &  & 334.225417 & 70.0625 & 374.5 & 6650 & 0.47 & 0.118 \\
BP Psc & 21814016 &  & 350.602917 & -2.228333 & 348.9 & 5350 & 0.73 & 0.828 \\
CO Ori & 21870336 &  & 81.909583 & 11.4275 & 404 & 6250 & 1.5 & 2.139 \\
CPM 25 & 25736192 &  & 95.984583 & 14.507778 & 2128.9 & 19500 & 2.85 & 3.825 \\
CQ Tau & 21875712 &  & 83.99375 & 24.748333 & 163.1 & 6750 & 0.87 & 0.406 \\
DG Cir & 16828160 & 16828160 & 225.849167 & -63.383056 & 832.9 & 11000 & 1.58 & 3.94 \\
DK Cha & 12679168 & 22349312 & 193.32125 & -77.119722 & 242.9 & 7250 & 0.47 & 8.12 \\
GSC 3975-0579 & 12683008 &  & 324.535417 & 57.446667 & 942.2 & 8900 & 1.53 & 0.778 \\
GSC 5360-1033 & 25735424 &  & 89.45625 & -14.092778 & 605.2 & 15000 & 1.01 & 1.6 \\
HBC 217 & 12675328 &  & 100.175833 & 9.560278 & 695.6 & 6250 & 0.79 & 0.062 \\
HBC 222 & 21880832 &  & 100.213333 & 9.746111 & 706.3 & 6250 & 0.82 & 0.115 \\
HBC 334 & 25731328 &  & 34.127917 & 55.383333 & 1774.4 & 16500 & 2.18 & 2.365 \\
HBC 442 & 18832640 &  & 83.559167 & -5.615 & 385.7 & 6170 & 0.98 & 0.074 \\
HBC 717 & 21885696 &  & 313.025 & 44.287778 & 1394.9 & 6400 & 1.88 & 2.824 \\
HD 101412 & 5640960 & 5640960 & 174.935 & -60.174444 & 411.3 & 9750 & 1.58 & 0.21 \\
HD 104237 & 12677632 & 12677632 & 180.020417 & -78.193056 & 108.4 & 8000 & 1.33 & 0 \\
HD 130437 &  & 16826368 & 222.709167 & -60.286111 & 1653.2 & 24500 & 4.31 & 2.61 \\
HD 132947 & 5643008 & 5643008 & 226.233333 & -63.131389 & 381.6 & 10250 & 1.61 & 0 \\
HD 135344B &  & 5657088 & 228.951667 & -37.154444 & 135.8 & 6375 & 0.79 & 0.23 \\
HD 141926 & 25739264 &  & 238.590833 & -55.328889 & 1338.2 & 28000 & 4.74 & 2.4 \\
HD 142527 & 11005696 & 11005696 & 239.174583 & -42.323333 & 157.3 & 6500 & 0.96 & 0 \\
HD 143006 & 5197568 & 9777152 & 239.65375 & -22.954444 & 166.1 & 5430 & 0.46 & 0.307 \\
HD 149914 & 11000832 & 11000832 & 249.619167 & -18.220556 & 158.8 & 10250 & 2.09 & 0.95 \\
HD 150193 &  & 11006208 & 250.074583 & -23.895833 & 150.8 & 9000 & 1.37 & 1.55 \\
HD 155448 & 11006464 & 11006464 & 258.245 & -32.242778 & 953.9 & 10700 & 2.74 & 0.468 \\
HD 163296 &  & 5650944 & 269.08875 & -21.956111 & 101.5 & 9250 & 1.2 & 0 \\
HD 17081 &  & 10998272 & 41.030417 & -13.858889 & 106.7 & 13000 & 2.58 & 0 \\
HD 174571 & 25740544 &  & 282.696667 & 8.702778 & 1095 & 19500 & 4.23 & 2.502 \\
HD 179218 &  & 11006976 & 287.797083 & 15.7875 & 266 & 9500 & 2.05 & 0.527 \\
HD 200775 & 16207104 & 16207104 & 315.40375 & 68.163333 & 360.8 & 16500 & 3.07 & 1.054 \\
HD 244604 & 11001344 & 11001344 & 82.98875 & 11.294722 & 420.6 & 9000 & 1.46 & 0.14 \\
HD 249879 & 25735680 &  & 89.7325 & 16.665833 & 669 & 11500 & 1.56 & 0.205 \\
HD 250550 & 16826624 & 16826624 & 90.5 & 16.515833 & 697.1 & 11000 & 1.94 & 0 \\
HD 259431 &  & 11003392 & 98.271667 & 10.322222 & 720.9 & 14000 & 2.97 & 1.11 \\
HD 288012 & 21889280 &  & 83.27 & 2.469444 & 395.8 & 9800 & 1.66 & 0.573 \\
HD 290380 & 21889024 &  & 80.879167 & -1.073333 & 354.3 & 6400 & 0.84 & 0.059 \\
HD 290764 & 21890304 &  & 84.522083 & -1.256111 & 397.9 & 7875 & 1.18 & 0.16 \\
HD 290770 & 25734656 &  & 84.26 & -1.6225 & 399.1 & 10500 & 1.52 & 0 \\
HD 319896 & 25739776 &  & 262.774583 & -35.141389 & 1295.2 & 15750 & 3.19 & 2.378 \\
HD 35929 & 10998528 & 10998528 & 81.928333 & -8.3275 & 387.4 & 7000 & 1.79 & 0 \\
HD 36112 & 11001088 & 11001088 & 82.614583 & 25.3325 & 160.3 & 7605 & 1.04 & 0.155 \\
HD 36408 & 25734400 &  & 83.05875 & 17.058056 & 435.1 & 11933 & 3.13 & 0.378 \\
HD 36917 & 11001600 & 11001600 & 83.695833 & -5.570833 & 474 & 11215 & 2.43 & 0.521 \\
HD 37258 & 18814720 & 10998784 & 84.247083 & -6.154444 & 362.7 & 9750 & 1.24 & 0.06 \\
HD 37357 & 11001856 & 11001856 & 84.44625 & -6.708333 & 649.6 & 9500 & 2.04 & 0 \\
HD 37806 & 11002368 & 11002368 & 85.259583 & -2.716944 & 427.6 & 10475 & 2.17 & 0.133 \\
HD 38087 & 11002624 & 11002624 & 85.7525 & -2.3125 & 338.1 & 13600 & 2.19 & 0.456 \\
HD 38120 & 11002880 & 11002880 & 85.799583 & -4.997222 & 405 & 10700 & 1.72 & 0.21 \\
HD 39014 &  & 10999296 & 86.192917 & -65.735556 & 44.1 & 7830 & 1.42 & 0 \\
HD 46060 & 25732864 &  & 97.7075 & -9.654167 & 932.9 & 21050 & 3.89 & 1.795 \\
HD 50083 & 11000064 & 11000064 & 102.940833 & 5.084444 & 1089.8 & 16500 & 4.04 & 0.676 \\
HD 50138 &  & 11003648 & 102.889167 & -6.966389 & 379.9 & 9450 & 2.46 & 0.031 \\
HD 53367 & 16826880 & 16826880 & 106.10625 & -10.454444 & 129.7 & 29500 & 3.13 & 2.051 \\
HD 56895B & 11003904 & 11003904 & 109.6325 & -11.192778 & 165.3 & 7000 & 0.97 & 0.081 \\
HD 58647 & 11004160 & 11004160 & 111.48375 & -14.178889 & 318.5 & 10500 & 2.44 & 0.372 \\
HD 59319 & 25736704 &  & 112.153333 & -21.963611 & 668.4 & 12500 & 2.51 & 0 \\
HD 72106B & 11004416 & 11004416 & 127.395417 & -38.605833 & 597.2 & 8750 & 1.85 & 0.505 \\
HD 76534 & 16827136 & 16827136 & 133.78625 & -43.466667 & 910.6 & 19000 & 3.55 & 0.62 \\
HD 85567 & 11004672 & 11004672 & 147.61875 & -60.9675 & 1023 & 13000 & 3.19 & 0.89 \\
HD 95881 & 11004928 & 11004928 & 165.49 & -71.513333 & 1168.3 & 10000 & 2.85 & 0 \\
HD 96042 & 14206464 &  & 165.91875 & -59.433056 & 3100.2 & 25500 & 4.81 & 0.78 \\
HD 9672 & 4928768 & 4928768 & 23.657917 & -15.676389 & 57.1 & 8900 & 1.17 & 0 \\
HD 97048 & 12697088 & 12697088 & 167.013333 & -77.654722 & 184.8 & 10500 & 1.54 & 0.9 \\
HD 98922 &  & 5640704 & 170.632083 & -53.369722 & 688.8 & 10500 & 3.03 & 0.09 \\
Hen 2-80 &  & 25738496 & 185.596667 & -63.288056 & 753.5 & 14000 & 2.12 & 2.967 \\
Hen 3-1191 &  & 16828928 & 246.812917 & -48.6575 & 1661.5 & 29000 & 3.49 & 3.841 \\
Hen 3-847 &  & 25733120 & 195.324167 & -48.888611 & 784.8 & 14000 & 2.07 & 0.57 \\
HR 5999 &  & 11005952 & 242.142917 & -39.105278 & 161.1 & 8500 & 1.72 & 0.33 \\
IL Cep & 25734144 &  & 343.315 & 62.145833 & 805.2 & 16500 & 3.86 & 3.159 \\
KK Oph &  & 16827392 & 257.53375 & -27.255278 & 221.1 & 8500 & 0.71 & 2.7 \\
LkHa 215 & 14124032 & 14124032 & 98.174167 & 10.159444 & 713.1 & 14000 & 2.57 & 2.018 \\
LkHa 257 & 25733888 &  & 328.578333 & 47.202778 & 793.8 & 15000 & 1.91 & 2.654 \\
LKHa 260 & 25747456 &  & 274.789167 & -13.844722 & 1234.5 & 14000 & 2.09 & 3.196 \\
LKHa 338 & 21880064 &  & 92.69625 & -6.214167 & 884.6 & 10700 & 1.13 & 2.601 \\
LkHa 339 & 25732352 &  & 92.740833 & -6.244444 & 857.1 & 10500 & 1.92 & 3.54 \\
MWC 1080 & 21887488 &  & 349.356667 & 60.845278 & 1203 & 29000 & 4.52 & 5.028 \\
MWC 137 &  & 26897920 & 94.689583 & 15.281111 & 2907.4 & 29000 & 4.94 & 4.634 \\
MWC 314 & 27569664 &  & 290.391667 & 14.8825 & 2977 & 16500 & 5.29 & 4.5 \\
MWC 593 & 25746432 &  & 267.2925 & -24.239167 & 1341.4 & 15750 & 3.58 & 2.313 \\
MWC 623 &  & 22902016 & 299.13125 & 31.105556 & 3279.8 & 15825 & 4.58 & 3.77 \\
MWC 657 & 22903552 & 22903552 & 340.674167 & 60.400278 & 3164.2 & 19850 & 4.62 & 5.031 \\
MWC 878 &  & 25739520 & 261.18625 & -38.730833 & 1773.8 & 24500 & 4.32 & 3.06 \\
MWC 930 & 25444352 &  & 276.605 & -7.221667 & 2585.8 & 11900 & 5.85 & 8.717 \\
NV Ori & 21876480 &  & 83.880833 & -5.5525 & 386.5 & 6750 & 1.19 & 0.127 \\
NX Pup & 21882112 & 16827904 & 109.867917 & -44.586389 & 1672.5 & 7000 & 2.46 & 0 \\
PDS 022 & 25735936 &  & 90.904583 & -14.884167 & 854.6 & 9800 & 1.82 & 0.205 \\
PDS 144S & 25680384 & 25680384 & 237.31375 & -26.015278 & 149.6 & 7750 & -0.67 & 0.57 \\
PDS 211 & 22894848 & 22894848 & 92.572083 & 29.421389 & 1073.8 & 10700 & 1.79 & 2.985 \\
PDS 241 &  & 25745408 & 107.161667 & -4.318056 & 2887.9 & 26000 & 4.05 & 2.6 \\
PDS 27 &  & 25736448 & 109.899583 & -17.655 & 2552.6 & 17500 & 4.15 & 5.03 \\
PDS 344 & 25738241 &  & 175.136667 & -64.535 & 2439.5 & 15250 & 2.24 & 0.86 \\
PDS 37 & 25447169 & 25737473 & 152.50125 & -57.035278 & 1925.5 & 17500 & 4 & 5.81 \\
PDS 415N & 12674304 & 12674304 & 244.655 & -24.088333 & 144.2 & 6250 & 0.44 & 1.476 \\
PDS 477 & 25740288 &  & 270.12625 & -16.790556 & 2471.9 & 24500 & 3.61 & 4.281 \\
PDS 543 & 25746688 & 25746688 & 282.002917 & 2.904722 & 1413.2 & 29000 & 5.21 & 7.121 \\
PDS 581 &  & 25740800 & 294.07875 & 29.547222 & 687.9 & 24500 & 2.89 & 2.626 \\
PDS 69 &  & 25739008 & 209.432917 & -39.979722 & 642.5 & 15000 & 2.7 & 1.6 \\
PX Vul & 21884928 &  & 291.667917 & 23.8975 & 627.4 & 6750 & 1.36 & 1.212 \\
RR Tau & 5638400 & 5638400 & 84.877083 & 26.374167 & 773.4 & 10000 & 2.01 & 1.55 \\
RY Ori & 21871616 &  & 83.04125 & -2.829722 & 368.5 & 6250 & 0.86 & 0.961 \\
SAO 185668 & 11006720 & 11006720 & 265.981667 & -22.095833 & 1481.6 & 16500 & 3.8 & 2.015 \\
SAO 220669 & 25737216 &  & 133.94125 & -44.420556 & 932.1 & 16000 & 3.58 & 1.89 \\
V1012 Ori & 25731584 &  & 77.902083 & -2.38 & 386.4 & 8500 & 0.77 & 1.32 \\
V1295 Aql & 11007232 & 11007232 & 300.760417 & 5.738056 & 870.9 & 9500 & 2.9 & 0.403 \\
V1478 Cyg &  & 7464704 & 308.189583 & 40.660278 & 1296.7 & 19850 & 5.17 & 8.909 \\
V1686 Cyg &  & 16827648 & 305.122083 & 41.357778 & 1078.8 & 6010 & 1.53 & 1.854 \\
V1787 Ori & 18834176 &  & 84.53875 & -6.821389 & 391 & 8150 & 1.15 & 4.076 \\
V1818 Ori &  & 25735168 & 88.4275 & -10.400278 & 695 & 13000 & 2.96 & 3.717 \\
V1977 Cyg & 21885184 &  & 311.90625 & 43.790278 & 859.8 & 11000 & 2.48 & 2.263 \\
V346 Ori & 16265216 &  & 81.178333 & 1.73 & 366.4 & 7750 & 0.92 & 0 \\
V373 Cep &  & 25733632 & 325.778333 & 66.115 & 922.1 & 13000 & 2.29 & 3.066 \\
V380 Ori &  & 25731840 & 84.105833 & -6.716111 & 481.7 & 9750 & 2 & 2.21 \\
V388 Vel & 18596864 & 18596864 & 130.572083 & -40.736111 & 2466.9 & 9500 & 2.45 & 3.996 \\
V590 Mon & 12674816 & 12674816 & 100.185833 & 9.800556 & 818.4 & 12500 & 1.38 & 1.03 \\
V594 Cas &  & 25731072 & 10.82625 & 61.911111 & 569.2 & 11500 & 2.13 & 1.9 \\
V669 Cep &  & 22903040 & 336.66125 & 61.225556 & 977.6 & 15000 & 2.6 & 3.047 \\
V892 Tau &  & 3869696 & 64.669167 & 28.320833 & 117.5 & 11500 & 0.13 & 4.867 \\
V921 Sco &  & 4898048 & 254.778333 & -42.702222 & 1545.6 & 29000 & 4.76 & 4.879 \\
VV Ser & 5651200 & 5651200 & 277.199583 & 0.144444 & 419.7 & 13800 & 1.95 & 2.908 \\
WW Vul & 16828672 & 16828672 & 291.495 & 21.208611 & 503.5 & 8970 & 1.42 & 0.949 \\ \hline
\end{longtable}

%%%%%%%%%%%%%%%%%%%%%%%%%%%%%%%%%%%%%%%%%%%%%%%%%%

\twocolumn

\section{A Brief description of the sources with C\textsubscript{60} features}

\subsection*{BD+30 549}

BD+30 549 is a member of the young proto-cluster NGC 1333, in the Perseus OB2 association \citep{Cernis1990Ap&SS.166..315C,Azimlu2015AJ....150...95A}. The optical image of the nearby regions of the star suggests that the star is associated with nebulosity \citep{Boersma2016ApJ...832...51B}. Also, intense PAH emission features are detected in the IRS spectrum of BD+30 549. The association of BD+30 549 with a young cluster, its young age, the presence of nebulosity in the vicinity and PAH emission features in the spectrum confirm the young PMS nature of the object.

There are multiple IRS spectra (both high and low resolution) pointed on the RNe associated with BD+30 549. The spatial evolution of PAH features around the RNe are discussed by \cite{Andrews2015ApJ...807...99A} and \cite{Boersma2016ApJ...832...51B}. We identified C\textsubscript{60} features in spectra pointed spatially away from the star. A study on the spatial evolution of C\textsubscript{60} features near BD+30 549 is underway (Arun et al. in prep).

\subsection*{PDS 216 (CPM 25)}

PDS 216 is reportedly near Lynd 1600 cloud (L1600; \citealp{Vieira2011A&A...526A..24V}). PDS 216 has been proposed  distinctively as a HAeBe star \citep{VIEIRA2003, Vioque2018A&A...620A.128V}, Asymptotic Giant Branch (AGB) star \citep{Vieira2011A&A...526A..24V} and a possible planetary nebulae (PNe; \citealp{Viironen2009A&A...504..291V}). The {\it WISE} classification scheme by \cite{Marton2016MNRAS.458.3479M} reports the star as a YSO. More importantly, the star is associated with an RNe named GN 06.21.1 \citep{Magakian2003A&A...399..141M}.

We evaluated the nature of PDS 216 from the {\it Spitzer} IRS spectrum analysis. In general, the emission lines belonging to [Ar II] 6.99 $\mu m$, [S IV] 10.51 $\mu m$, [NeII] 12.81 $\mu m$ can be seen in the MIR spectrum of a PNe \citep{ramos-2016}. None of these forbidden lines are found in the IRS spectrum of PDS 216. Also, intense PAH emission features are identified in the spectrum. This suggests that PDS 216 is a HAeBe star.

\subsection*{HBC 334}

HBC 334 is associated with the reflection nebulae RNO 6 \citep{Magakian2003A&A...399..141M}. The clustering of stars around this HAeBe star is studied by \cite{Testi1998}. The IRS spectrum shows intense PAH emission features at 6.2, 7.7, 8.6, 11.2 and 12.4 $\mu m$. The association of RNe, clustering properties, and the PAH in the IRS spectrum shows that HBC 334 is a HAeBe star.

\subsection*{HD 319896}

HD 319896 is near the ``twin star-forming complex" NGC 6357 and NGC 6334 \citep{Russeil2012A&A...538A.142R}. HD 319896 is reported to show a more evolved nature. \cite{Szczerba2007A&A...469..799S} reported the star as a candidate AGB whereas \cite{Verhoeff2012A&A...538A.101V}, using MIR imaging, proposed that the star might be a Classical Be star due to its low IR excess.

We downloaded the optical spectrum of HD 319896 from the ESO archive, which was observed using the HARPS spectrograph \citep{Pepe2000SPIE.4008..582P, Mayor2003Msngr.114...20M}. The H$\alpha$ emission line of HD 319896 shows a double peak profile, suggesting the association of a circumstellar disc with the star. Thus, the possibility of HD 319896 being an AGB can be ruled out. Also, our analysis showed that HD 319896 is a Class II source (\autoref{sec:index}). Classical Be stars do not show a flat SED. Hence, by analysing the available data, we conclude that HD 319896 is a HAeBe star.

\subsection*{PDS 344 (BRAN 366)}

PDS 344 is associated with a reflection nebulae RNO 555 \citep{Magakian2003A&A...399..141M}. \cite{FAirlamb2015} modelled the magnetospheric accretion of the star using the X-shooter spectrum from the evaluation of the ultraviolet excess. They estimated a mass accretion rate of log($\dot{M}_{acc}$) = -7.02 for PDS 344 using the excess values, which is consistent with the $\dot{M}_{acc}$ values of HAeBe stars of similar mass. 

\subsection*{SAO 220669}

The X-shooter spectrum of SAO 220669 show no emission lines. However, using the spectral information, \cite{VIEIRA2003} subtracted the absorption component of H$\alpha$ and found underlying emission ($\sim$ 1 $nm$). The star is also identified as a Class II source. Based on the H$\alpha$ emission and the infrared excess, it can be concluded that SAO 220669 is a HAeBe star.

\subsection*{LkH$\alpha$ 215}

LkH$\alpha$ 215 is associated with RNe NGC 2245. The IRS spectra of show PAH emission features and 17 $\mu m$ plateau along with 17.4 and 18.9 $\mu m$ C\textsubscript{60} features. The spectra is similar to that of a HAeBe stars associated with nebulosity. The spectra are shown in \autoref{fig:fullrnelike}.

\subsection*{MWC 593}
Various studies classified MWC 593 as a HAeBe star \citep{The1994,Guzm2021A&A...650A.182G}. \cite{Verhoeff2012A&A...538A.101V}, using MIR imaging, found that 12 $\mu m$ IR excess but is very faint, which is consistent with in the 5-15 $\mu m$ region of IRS spectra, which has lower flux. 

\subsection*{HD 46060}

HD 46060 is located near the Southern Filament of Orion and Mon R2 \citep{Kim2004}. The star is associated with a reflection nebulae RNO 59 \citep{Magakian2003A&A...399..141M}. The {\it WISE} classification scheme by \cite{Marton2016MNRAS.458.3479M} reports the star as a YSO.  
% Don't change these lines
\bsp	% typesetting comment
\label{lastpage}
\end{document}